\documentclass[journal]{IEEEtran}
\IEEEoverridecommandlockouts

\usepackage{amsmath,amssymb,amsfonts}
\usepackage{algorithmic}
\usepackage{textcomp}
\usepackage[table,xcdraw]{xcolor}
\usepackage[nolist]{acronym}
\usepackage{xurl}
\usepackage{paralist}
\usepackage{multirow}
\usepackage{listings}
\usepackage{tabularx}
\usepackage{rotating}
\usepackage{caption}
\usepackage{subcaption}
\usepackage{graphicx}
\usepackage{hhline}
\usepackage{arydshln}
\usepackage{tikz,float}
\usepackage{array}
\usepackage{xparse}
\usepackage{tcolorbox}
\usepackage{xcolor}
\usepackage{xspace}
\usepackage{pgfplots, pgfplotstable}
\usepackage{siunitx,  
            booktabs, 
            caption}  
\usepackage{xcolor}
\usepackage[switch]{lineno}
\usepackage{comment}
\usepackage{cite}
\usepackage{adjustbox}
\usepackage{pdflscape}
\usepackage{hyperref}
\usepackage{tikz}

\colorlet{NextBlue}{red!25!green!50!blue!75}

\tikzset{
  font={\fontsize{10pt}{12}\selectfont}}
\usetikzlibrary{trees}
\usetikzlibrary{patterns, backgrounds}

\definecolor{ao(english)}{rgb}{0.0, 0.5, 0.0}
\definecolor{bananayellow}{rgb}{1.0, 0.88, 0.21}
\definecolor{bostonuniversityred}{rgb}{0.8, 0.0, 0.0}
\definecolor{brandeisblue}{rgb}{0.0, 0.44, 1.0}
\usepackage{forest}
\tikzset{
  treenode/.style={
      shape=rectangle,
      rounded corners,
      ultra thick,
      draw=brandeisblue,
      align=left,
      fill=white
  }
}

  \def\mybarc#1{
  {\rotatebox{90}{\color{gray}\rule{#1pt}{4pt}}}}

\NewDocumentCommand{\statcirc}{ O{#2} m }{%
    \begin{tikzpicture}
    \fill[#2] (0,0) circle (1.0ex); 
    \fill[#1] (0,0) -- (90:1ex) arc (90:330:1ex) -- cycle; 
    \end{tikzpicture}
}

\definecolor{codegreen}{rgb}{0,0.6,0}
\definecolor{codegray}{rgb}{0.5,0.5,0.5}
\definecolor{codepurple}{rgb}{0.58,0,0.82}
\definecolor{backcolour}{rgb}{0.95,0.95,0.92}

\lstdefinestyle{mystyle}{
    backgroundcolor=\color{backcolour},   
    commentstyle=\color{codegreen},
    keywordstyle=\color{magenta},
    numberstyle=\tiny\color{codegray},
    stringstyle=\color{codepurple},
    basicstyle=\ttfamily\footnotesize,
    breakatwhitespace=false,         
    breaklines=true,                 
    captionpos=b,                    
    keepspaces=true,                 
    numbers=left,                    
    numbersep=5pt,                  
    showspaces=false,                
    showstringspaces=false,
    showtabs=false,                  
    tabsize=2
}

\newcommand{\mm}[1]{}
\newcommand{\ob}[1]{}
\newcommand{\hp}[1]{}

\def\BibTeX{{\rm B\kern-.05em{\sc i\kern-.025em b}\kern-.08em
    T\kern-.1667em\lower.7ex\hbox{E}\kern-.125emX}}
\begin{document}

\title{Taxonomy of Attacks on Open-Source Software Supply Chains}


\author{

\IEEEauthorblockN{Piergiorgio Ladisa\IEEEauthorrefmark{1}\IEEEauthorrefmark{3},
Henrik Plate\IEEEauthorrefmark{1},
Matias Martinez\IEEEauthorrefmark{2}, and 
Olivier Barais\IEEEauthorrefmark{3},}

\IEEEauthorblockA{\IEEEauthorrefmark{1}SAP Security Research}
\IEEEauthorblockA{\IEEEauthorrefmark{2}Universit\'{e} Polytechnique Hauts-de-France}
\IEEEauthorblockA{\IEEEauthorrefmark{3}Universit\'{e} de Rennes 1, Inria, IRISA}
\IEEEauthorblockA{\{piergiorgio.ladisa, henrik.plate\}@sap.com, matias.martinez@uphf.fr, \{piergiorgio.ladisa, olivier.barais\}@irisa.fr}
}

\maketitle


\begin{abstract}
    The widespread dependency on open-source software makes it a fruitful target
    for malicious actors, as demonstrated by recurring attacks. The
    complexity of today's open-source supply chains results in a significant
    attack surface, giving attackers numerous opportunities to reach the goal of
    injecting malicious code into open-source artifacts that is then downloaded
    and executed by victims.
    
    This work proposes a general taxonomy for attacks on open-source supply
    chains, independent of specific programming languages or ecosystems, and
    covering all supply chain stages from code contributions to package
    distribution. Taking the form of an attack tree, it covers 107 unique
    vectors, linked to 94 real-world incidents, and mapped to 33 mitigating
    safeguards.
    
    User surveys conducted with 17 domain experts and 134 software developers
    positively validated the correctness, comprehensiveness and
    comprehensibility of the taxonomy, as well as its suitability for various
    use-cases. Survey participants also assessed the utility and costs of the
    identified safeguards, and whether they are used.
\end{abstract}


\section{Introduction}\label{section:intro}



\textit{Software supply chain attacks} aim at injecting malicious code into
software components to compromise downstream users. Recent
incidents, like the infection of SolarWind's Orion platform~\cite{9382367},
downloaded by approx. 18,000 customers, including government agencies and
providers of critical infrastructure, demonstrate the reach and potential impact
of such attacks. Accordingly, software supply chain attacks are among the
primary threats in today's threat landscape, as reported by
ENISA~\cite{ENISAThreat}
or the 
\textit{US Executive Order on Improving the Nation’s
Cybersecurity}~\cite{executiveorder}.

This work focuses on the specific instance of attacks on \ac{OSS} supply chains,
which exploit the widespread use of open-source during the software
development lifecycle as a means for spreading malware. Considering the
dependency of the software industry on open-source \textendash~across the technology stack
and throughout the development lifecycle, from libraries and frameworks to
development, test and build tools, Ken Thompson's
reflections~\cite{10.1145/358198.358210} on trust (in code and its authors) is
more 
relevant than ever. Indeed, attackers abuse trust relationships existing between
the different open-source stakeholders~\cite{herr2021breaking,chess2007attacking}.
The appearance and significant increase of attacks on \ac{OSS} throughout the
last few years, as reported by Sonatype in their 2021
report~\cite{sonatypereport}, demonstrate that attackers consider them a viable
means for spreading malware.


Recently, industry and government agencies increased their efforts to improve
software supply chain security, both in general and in regards to open-source.
MITRE, for instance, proposes an end-to-end framework to preserve supply chain
integrity~\cite{deliveruncompromised}, and the OpenSSF develops the SLSA
framework, which groups several security best-practices for open-source
projects~\cite{slsaframework}. Academia contributes an increasing number of
scientific publications, many of which get broad attention in the developer
community, e.g., ~\cite{boucher2021trojan} or \cite{wu2021feasibility}.


Nevertheless, we observed that existing works on open-source supply chain
security lack a comprehensive, comprehensible, and general description of how
attackers inject malicious code into \ac{OSS} projects, that is independent of
specific programming languages, ecosystems, technologies, and stakeholders.


We believe a taxonomy classifying such attacks could be of value for both
academia and industry. Serving as a common reference and clarifying terminology,
it could support several activities, e.g., developer training, risk assessment,
or the development of new safeguards. As such, we set out to answer the
following research questions:



\noindent\textbf{RQ1 \textendash~Taxonomy of attacks on \ac{OSS} supply chains}
\begin{itemize}
    \item RQ1.1 \textendash~What is a comprehensive list of general attack vectors on
    \ac{OSS} supply chains?
    \item RQ1.2 \textendash~How to represent those attack vectors in a comprehensible and
    useful fashion?
\end{itemize}

\noindent\textbf{RQ2 \textendash~Safeguards against \ac{OSS} supply chain attacks}
\begin{itemize}
    \item RQ2.1 \textendash~Which general safeguards exist, and which attack
    vectors do they address?
    \item RQ2.2 \textendash~What is the utility and cost of those safeguards?
    \item RQ2.3 \textendash~Which safeguards are used by developers?
\end{itemize}



To answer those questions, we first study both the scientific and grey
literature to compile an extensive list of attack vectors, including ones that
have been exploited, but also non-exploited vulnerabilities and plausible
proofs-of-concept.
We then outline a taxonomy in the form of an attack tree. From the identified
attacks, we list the associated safeguards. Finally, we conduct two user surveys
aiming to validate the attack taxonomy and to collect qualitative feedback
regarding the utility, costs, awareness, and use of safeguards.



To this extent, the main contributions of our work are as follows:
\begin{itemize}
    \item A taxonomy of \textbf{107 unique attack vectors} related to \ac{OSS}
    supply chains, taking the form of an attack tree and \textbf{validated by 17
    domain experts} in terms of completeness, comprehensibility, and
    applicability in different use cases. 
    \item A set of \textbf{33 safeguards} geared towards the proposed taxonomy,
    and qualitatively \textbf{assessed regarding utility and costs} by the same
    17 domain experts.
    \item The qualitative assessment of \textbf{134 developers} on the awareness
    of selected high-level attack vectors and the corresponding level of
    protection.
\end{itemize}

Using an interactive visualization of the attack tree, the taxonomy with
descriptions, examples of real-world attacks, references, and associated
safeguards can be explored online\footnote{\url{https://sap.github.io/risk-explorer-for-software-supply-chains/}}.


The remainder of the paper is organized as follows.
Section~\ref{section:background} introduces basic concepts and elements of
\ac{OSS} supply chains, the assumed attacker model, and the concept of attack
trees. Section~\ref{section:methodology} describes the methodology applied,
comprising the three steps \ac{SLR}, modeling of taxonomy and safeguards, and
survey design. Section~\ref{section:attacktaxonomy} details the proposed
taxonomy and presents the results of the expert and developer validation.
Section~\ref{section:safeguards} introduces the safeguards associated with the
aforementioned attack vectors and presents both the experts' feedback on their
utility and costs, and the developers' feedback on awareness and use. 
Section~\ref{section:discussion} discusses the differences between programming languages 
and highlights the benefits of our work on research.
Section~\ref{section:usersurveydemographics} provides demographic information
about the survey participants. Section~\ref{section:relatedworks} mentions
related works, and Section~\ref{section:threatovalidity} discusses threats to
the validity of our work. Finally, the conclusion and outlook are provided in
Section~\ref{section:conclusion}.

\section{Background}\label{section:background}

\begin{figure}[htp]
    \centering
    \includegraphics[width=.5\textwidth]{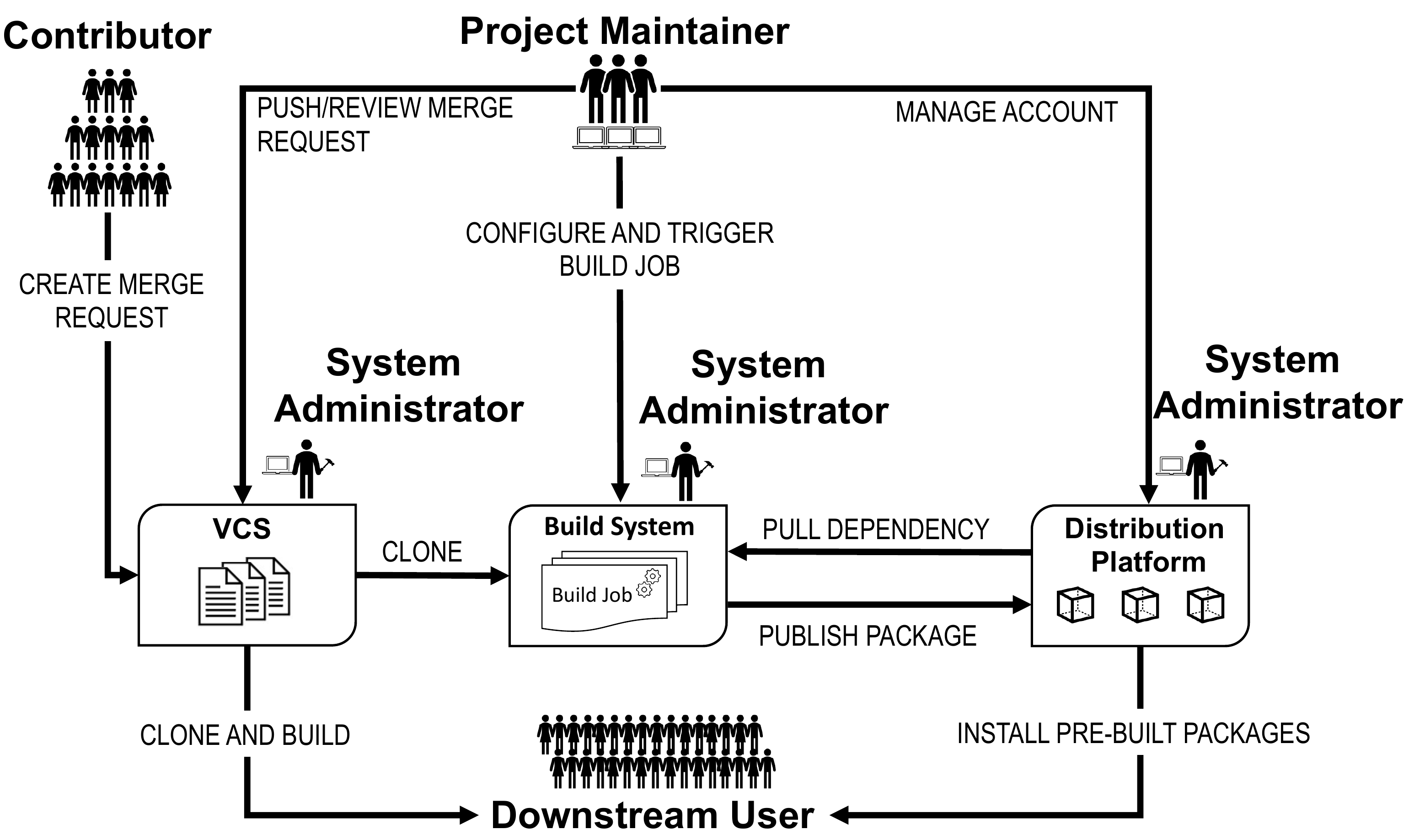}
    \caption{Stakeholders, systems and dataflows \hp{?} related to
    the development, build and distribution of \ac{OSS} artifacts.}
    \label{fig:sdlc}
\end{figure}

This section describes, at a high level, the systems and stakeholders involved
in the development, build, and distribution of \ac{OSS} artifacts (cf.
Figure~\ref{fig:sdlc}). They are constituting elements of \ac{OSS} supply chains
and contribute to their attack surface. They commonly interact in a distributed
setting~\cite{haddad2011understanding}, even if the specifics differ from one
\ac{OSS} project to another. 

The section concludes with a description of the attacker model considered
throughout the paper, and a summary of the concept of attack trees.

\subsection{Systems}
The systems considered comprise \ac{VCS}, build systems, and package
repositories. They do not necessarily correspond to concrete physical or virtual
systems providing the respective function but should be seen as roles,
multiple of which can be exercised by a single host or 3rd party service.

\textbf{\acl{VCS}s}
host the source code of the \ac{OSS} project, not only program code but also
metadata, build configuration, and other resources. They track and manage all
the changes of the codebase that happen throughout the development process.
Plain \ac{VCS}s like Git do not require its users to authenticate, but
complementary tools and 3rd party services offer additional functionalities
(e.g., issue trackers) or security controls (e.g., authentication, fine-grained
permissions or review workflows).


\textbf{Build Systems}
take a project's codebase as input and produce a binary artifact, e.g., an
executable or compressed archive, which can be distributed to downstream users
for easy consumption. The build commonly involves so-called dependency or
package managers~\cite{cox2019our,packagemanagementsecurity}, e.g., pip for
Python, which determine and download all dependencies necessary for the build to
succeed, e.g., test frameworks or \ac{OSS} libraries integrated into the project
at hand. \ac{CI}/\ac{CD} pipelines, running on build automation tools like
Jenkins, automate the test, build, and deployment of project artifacts.


\textbf{Distribution Platforms}
distribute pre-build \ac{OSS} artifacts to downstream users, e.g.,
upon the execution of package managers or through manual download. Our
definition does not only cover well-known public package repositories like PyPI
or Maven Central but also internal and external mirrors, \ac{CDN} or proxies. 


\textbf{Workstations of \ac{OSS} Maintainers and Administrators.}
\ac{OSS} project maintainers and administrators of the abovementioned systems
have privileged access to sensitive resources, e.g., the codebase, a build
system's web interface, or a package repository's database. Therefore, their
workstations are in the scope of the attack scenario.

\subsection{Stakeholders}
The stakeholders considered comprise \ac{OSS} project maintainers, contributors, and
consumers \textendash~as well as administrators of various systems or services involved.
Again, they should be understood as roles~\cite{duan2020towards}, multiple of
which can be assumed by a given individual. For example, maintainers of an
\ac{OSS} project typically consume artifacts of other projects.

\textbf{Contributors}
contribute code to an \ac{OSS} project, with limited (read-only)
access to project resources. They typically submit contributions to the \ac{VCS}
via merge requests, which are reviewed by project maintainers prior to
being integrated. 

\textbf{\ac{OSS} Project Maintainers}
\label{sec:bg:maintainers}
have privileged access to project resources, e.g., to review and integrate
contributors' merge requests, configure build systems and trigger build jobs, or
deploy ready-made artifacts on package repositories.
The real names of project collaborators, both contributors and maintainers, are
not necessarily known. Accounts, including anonymous ones, gain trust through
continued contributions of quality, thanks to which they may be promoted to
maintainers (known as a meritocracy).

\textbf{System and Service Administrators}
\label{sec:bg:admin}
have the responsibility to configure, maintain, and operate any of the
above-mentioned systems or services, e.g., employees of 3rd-party Git hosting
providers, members of \ac{OSS} foundations that operate own build systems for
their projects, or employees of companies running package repositories like npm.


\textbf{Downstream Users}
consume \ac{OSS} project artifacts, e.g., its source code from the project's
\ac{VCS}s (e.g., through cloning), or pre-built packages from distribution
platforms. In the context of downstream development projects, the download is
typically automated by package managers like pip or npm, which
identify and obtain dozens or hundreds~\cite{236368,dann2021identifying} of 
the project's direct and transitive dependencies.

\subsection{Risks of Open-Source Software Supply Chains}

OSS is widely used by organizations and individuals across the technology stack
and throughout the software development lifecycle. Package managers automate its
download and installation to a great extent, e.g., when resolving transitive
dependencies or updating versions.

The above-described systems are inherently distributed, and the stakeholders are
partly unknown or anonymous. They exist for every single open-source component
used, which multiplies an attack surface having both technical and social
facets. Moreover, even heavily used open-source projects receive only little
funding and contributions~\cite{10.1145/2663716.2663755}, making it difficult
for maintainers to securely run projects and increasing their susceptibility to
social-engineering attacks, e.g., when reviewing contributions.

Downstream consumers have no control over and limited visibility into given
projects' security practices. The sheer number of 
dependencies~\cite{236368} makes rigorous reviews impractical for a
given consumer, forcing them to trust the community for a timely detection of
vulnerabilities and attacks.

Attackers' primary objectives are data exfiltration, droppers, denial of
service, or financial gain~\cite{ohm2020backstabbers}. Hence, the larger the
user base, direct and indirect, the more attractive an open-source project
becomes for attackers. As in other adversarial contexts, attackers require
finding single weaknesses, while defenders needs to cover the whole attack
surface, which in this case spans the whole supply chain.




\subsection{Attack Tree}




Attack trees~\cite{schneier1999attack,foundationsattacktrees} are intuitive and
systematic representations of attacker goals and techniques, and support
organizations in risk assessment, esp. with regards to understanding exposure
and identifying countermeasures.

The root node of an attack tree represents the attacker's top-level goal, which
is iteratively refined by its children into subgoals. Depending on the degree
of refinement, the leaves correspond to more or less concrete and actionable
tasks.


As taxonomies require assigning instances to exactly one class, we only consider
disjunctive refinement, where child nodes represent alternatives to reach the
parent goal.



\subsection{Attacker Model}
\label{section:attackermodel}

The development of the taxonomy was based on the following assumptions and
attacker model.

The attacker's top-level goal is to place malicious code in
open-source artifacts such that it is executed in the context of downstream
projects, e.g., during its development or runtime. Such malware can exfiltrate
data, represent or open a backdoor, as well as download and execute second-stage
payload (e.g., cryptominers~\cite{ohm2020backstabbers}). Targeted assets can
belong both to developers of downstream software projects, or their end-users,
depending on the attacker's specific intention. However, the focus of the
taxonomy is not on \textit{what} malicious code does, but \textit{how} attackers
place it in upstream projects.


Insider attacks are out of scope, i.e., adversaries are neither maintainers of
the attacked open-source project nor members or employees of 3rd party service
providers involved in the development, build, or distribution of project
artifacts. As such, attackers do not have any privileged access to project resources
like build jobs or infrastructure like the server or database underlying code
repositories.

Initially, they only have access to publicly available information and publicly
accessible resources, which they can collect and analyze following the
\ac{OSINT}~\cite{GLASSMAN2012673} approach. Of course, due to the nature of
open-source projects, many project details are freely accessible, e.g., project
dependencies, build information, or commit and merge request histories.
Attackers can interact with any of the stakeholders and resources depicted in
Figure~\ref{fig:sdlc}, e.g., to communicate with maintainers using merge requests
or issue trackers or to create fake accounts and projects.



\begin{figure}[htp]
    \centering
    \includegraphics[width=.5\textwidth]{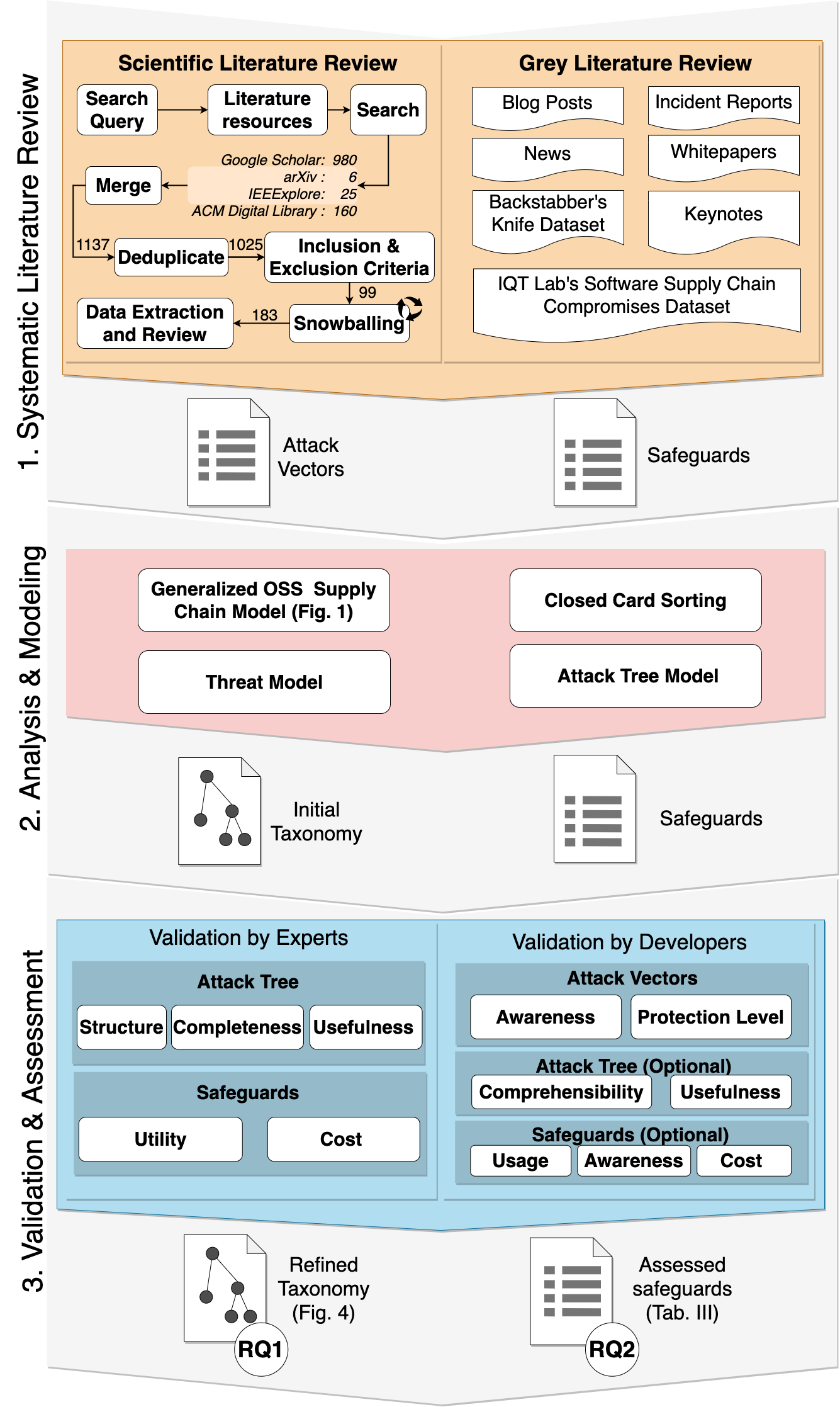}
    \caption{Our methodology comprises a literature review, the
    modeling of taxonomy \& safeguards, and the validation.}
    \label{fig:workflow}
\end{figure}

\section{Methodology}\label{section:methodology}

The methodology adopted to answer the above-mentioned research
questions comprises three phases (cf. Figure~\ref{fig:workflow}).

First, we review scientific and grey literature to collect an extensive list of
attack vectors on \ac{OSS} supply chains.

Second, starting from the vectors described in the literature and the \ac{OSS}
supply chain elements introduced in Section~\ref{section:background}, we
abstract from specific programming languages or ecosystems, perform threat
modeling, and create a taxonomy that takes the form of an attack tree. Also, we
identify and classify safeguards mitigating those vectors.

Third, to validate the proposed taxonomy and the list of safeguards, we design
and run two user surveys: with experts in the domain of \ac{OSS} supply chain
security, and with software developers, which are heavy consumers of \ac{OSS}.

\subsection{\acl{SLR}}
\label{section:slr}
The \ac{SLR} accomplishes two goals. First, through exploring the
state-of-the-art of \ac{OSS} supply chain security, we identify and
specify the abovementioned research questions. Second, it supports identifying
and collecting relevant attack vectors and suitable safeguards.
The \ac{SLR} itself follows a three step methodoloy comprising \textit{planning,
conducting}, and \textit{reporting}~\cite{kitchenham2009systematic}
\cite{wohlin2012experimentation} depicted in
Figure~\ref{fig:workflow} and described hereafter.


\subsubsection*{Search Strategy}
\label{section:searchstrategy}

This step defines the search terms, the query used on the identified resources,
and the inclusion criteria. For our purpose, we used the following query to search
for the terms anywhere in the documents:


\begin{lstlisting}[label={lst:query},numbers=none]
("open source" OR "open-source" OR "OSS" OR "free" 
OR "free/libre" OR "FLOSS) AND "software" AND 
("supply chain" OR "supply-chain") AND ("security" 
OR "insecurity" OR "attack" OR "threat" 
OR "vulnerability")
\end{lstlisting}

The four digital libraries used to collect the primary
studies are\footnote{Their URLs can be found in Appendix~\ref{appendix:C}}: Google Scholar
(980 results), arXiv (6), IEEExplore (25) and ACM Digital Library (160).
After removing duplicates from the total of 1171 search results, 1025 papers
remained.

We only included peer-reviewed articles in journals and conferences, technical
reports, and Ph.D./Master theses written in English and published before
March 2022.
Also, we only included studies related to security aspects, threats and malware
in the areas of \ac{OSS} development, \ac{VCS}, build systems and package
repositories, as well as malware detection and software supply chain security.
The application of those inclusion criteria reduced the 1025 results obtained in
the previous phase to 99 papers. 

We then applied the \textit{snowballing technique} on all those works to find
resources missed during the initial search, thereby applying the same inclusion
criteria. This resulted in the addition of another 84 new studies.

\subsubsection*{Data Extraction}
\begin{figure}[htbp]
    \centering
    \includegraphics[width=0.49\textwidth]{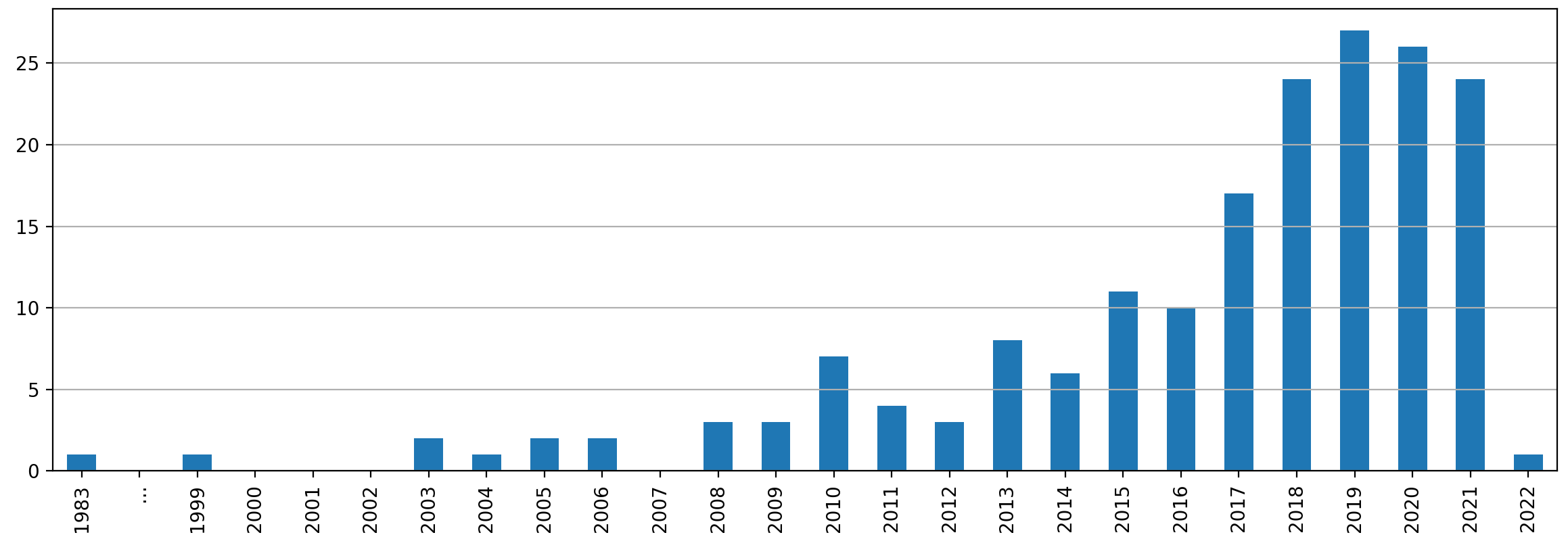}
    \caption{No. selected scientific articles per year of publication.}
    \label{fig:publicationsperyear}
\end{figure}

The selection process resulted in a total of 183 scientific works, mostly from
the last few years (cf.  Figure~\ref{fig:publicationsperyear}), which were
carefully reviewed to extract information about common threats, attack vectors,
and related safeguards. The complete list of the selected works is accessible
online~\footnote{\url{https://doi.org/10.5281/zenodo.6395965}}.




\subsection{Grey Literature}
\label{section:incidentreports}
\label{section:greyliterature}

In addition to scientific literature, especially to cover as many real-world attacks
and vulnerabilities as possible, we looked at grey literature like blog posts,
whitepapers, or incident reports. To this end, we periodically reviewed several
news aggregators and blogs (cf. Appendix~\ref{appendix:C}). Also, we used the
same search query as in Section~\ref{section:slr} for searching on Google. All
results were filtered using the selection criteria from
Section~\ref{section:slr}, and the \textit{snowballing technique} was applied to
further extend the set of sources.

\subsection{Analysis and Modeling of the Attack
Scenario}\label{section:threatmodeling} We perform the analysis of the \ac{OSS}
supply chain depicted in Figure~\ref{fig:sdlc} to classify the identified
attack vectors during the \ac{SLR}. Then we model such attacks using the
semantic of attack trees. The goal of these two steps is to answer to
\textbf{RQ1.2}, i.e., propose a taxonomy of \ac{OSS} supply chain attacks.

The analysis of the attack scenario in the context of \ac{OSS} development
started from the identification of the stakeholders (i.e., \textit{actors}),
\textit{systems}, as well as their relationship (i.e., \textit{channels}). We
have described such elements in Section~\ref{section:background} and depicted in
Figure~\ref{fig:sdlc}. This analysis was useful to identify potential categories
to structure the identified attack vectors. 

During the modeling phase, we adopted an \textit{attack-centric} methodology
whose purpose is to characterize the hostility of the environment and the attack
complexity for exploiting a system vulnerability~\cite{TUMA2018275}. In
particular, we performed closed card-sorting in the form of a
tree-test intending to build a taxonomy of \ac{OSS} supply chain
attacks as an attack tree. \text{Closed card-sorting} is an information
architecture technique taken from \ac{UX} design, in which the participants are
asked to structure a given set of information~\cite{spencer2009card}. A
tree-test is a particular case of a card-sorting problem, where the information
is structured in a tree.

For the attack tree modeling, we used as a starting point the attack tree
proposed by Ohm et al.~\cite{ohm2020backstabbers}, whose root node is
\textit{Injection of Malicious Code (into dependency tree)}. Thanks to a
rigorous structure, deeper refinement and the \ac{SLR}, we identified many
additional attack vectors (107 instead of 19). 

The main criteria to structure the attack tree were: degree of interference
with existing ecosystems (1st-level nodes), stages of the software supply chain
(i.e., source, build, distribute), and the system and stakeholders involved in
each stage.



The initial naming and arrangement have been changed to reflect the expert
feedback described in Section~\ref{sec:expertsurvey}. The refined version of our
initial attack tree is depicted in Figure~\ref{fig:attacktreetaxonomy}.

\subsection{Identification and Classification of Safeguards}
\label{section:safeguardsidentification}


To identify general safeguards, also in this case we reviewed the scientific
and grey literature described in
Section~\ref{section:greyliterature}. Then, each safeguard is classified
according to control type, stakeholder involvement, and mitigated attack
vector(s).


\textbf{Control type} classification follows the well-known distinction of directive,
preventive, detective, corrective, and recovery
controls~~\cite{wright2008regulatory}. However, since our focus is on \textit{how}
malicious code \textendash~no matter its actual intent \textendash~can be injected into
open-source and corresponding safeguards explains why recovery controls were
out of scope.


\textbf{Stakeholder involvement} reflects which role(s), maintainers, system
administrators or consumers, can or must become active to effectively implement a given
control.

Finally, each safeguard has been assigned to those node(s) of the attack tree
that it mitigates (partially or fully). To reflect the broader or narrower scope,
they were assigned to the tree node with the least possible depth.

\subsection{Survey Methodology} \label{section:surveymethodology}


We conducted two online surveys targeting two different audiences. First, we
addressed experts in the domain of software supply chain security to validate
the proposed taxonomy of attack vectors (\textbf{RQ1}) and to collect feedback
regarding the utility and costs of safeguards (\textbf{RQ2.2}).
Second, we addressed developers to rate their use of attack vectors and
perceived protection level. Optionally, they could additionally assess the taxonomy and
the use and awareness of safeguards from the perspective of open-source
consumers (\textbf{RQ2.3}). 


\subsubsection*{Questionnaire Design and Development}
We conducted a cross-sectional survey~\cite{blackstone2018principles} consisting
of the following four parts.

\paragraph*{Demographics}

This part collects background information about survey participants, especially
their skillset to check whether our objectives to address security experts and
developers are met, but also programming languages used, or whether they
actively participate in \ac{OSS} projects. The results are discussed in
Section~\ref{section:usersurveydemographics}.

\paragraph*{Taxonomy}\label{section:surveyTaxonomy}


In the expert survey, this part was meant to validate and assess the proposed
taxonomy. Before displaying our proposed taxonomy in its entirety, we used
tree-testing~\cite{albert2013measuring} to capture how easily
users find tree nodes. This helped validate the nodes' parent-child
relationships. Afterwards, participants were asked to explore an interactive
visualization of the complete taxonomy, and then to rate its structure, node
names, coverage, and its usefulness (to support different use-cases) on a Likert
scale from 1 (low) to 5 (high).


In the developer survey, this part started with a
presentation of the taxonomy's first-level nodes, including attack vector names and descriptions. Participants were asked whether they are aware of such attacks and whether they \textendash~or their organization \textendash~use any mitigating safeguards.
Optionally, participants could continue this part to explore the taxonomy and rate
its comprehensibility and usefulness.



\paragraph*{Safeguards}

In the expert survey, the participants assessed the utility and costs of the
selected safeguards. To this end, they were grouped by and presented according
to the stakeholder roles involved in their implementation.

This entire part was optional in the developer survey. When opting-in,
respondents only rated safeguards relevant according to their role in
open-source projects (if any). When shown, survey participants provided feedback
whether they use a given safeguard and its perceived costs (Likert scale). 

\subsubsection*{Pilot Survey and Pretest}
Interviews with two experts in user research and \ac{UX} provided us feedback on
the suitability and understandability of the survey.
Their main suggestions were to shorten texts and improve content presentation,
esp. of the tree-testing content.
After implementing their feedback, we performed a pretest of the expert survey
with 37 researchers from academia (i.e. Ph.D. students, researchers, and
professors), and the developer survey with 14 master students. The feedback
received from this pretest suggested further shortening texts, improving some
questions, and adjusting the appearance of buttons.

\subsubsection*{Sampling}


For the selection of participants in both questionnaires, we adopted the 
snowball sampling~\cite{10.1214/aoms/1177705148} technique. It consisted
of inviting an initial group of participants, which were asked to further share
the invitation in their appropriate network of knowledge. Due to this sampling
technique, it is not possible to compute the response rate.


The initial list of domain experts was composed of authors of works analyzed
during the \ac{SLR}, as well as experts from industry and academia from our
network. We included experts who performed relevant works in the context of
\ac{OSS} supply chain security (e.g, scientific publications,
initiatives/projects of software foundations or industry).
Similarly, the initial list of software developers has been created
starting from our network of knowledge of practicioners. 

The channels used to reach the participants have been emails, Linkedin, and direct
recruitment during presentations. 


The expert survey campaign began on 22 July 2021, the one for developers on
19 October 2021. Both questionnaires were closed for analysis on 24 November
2021, and reached a total of 17 and 134 respondents respectively.


\subsubsection*{Survey Procedure and Data Protection}

Rather than using existing survey tools or services, we developed a custom
solution. SurveyJS\footnote{\url{https://surveyjs.io/}} was used to design the
survey structure and content, which we exported as JSON file. This file was
hosted using GitHub Pages, together with SurveyJS' runtime library and other
resources. Participant answers were sent to a custom Google AppScript, which
stored them in a Google spreadsheet. Answers were sent after each survey page
and grouped using a random number generated in the beginning.


Applying the principle of data minimization, we did not collect IP addresses,
names or other PII. We also did not have access to 3rd party server logs.
Moreover, the decoupling of survey frontend and backend made that the first 3rd
party service provider only knows survey structure and content, while the second
only sees (encoded) answers without understanding their semantics. 


\section{Attack Taxonomy and its Validation}
\label{section:attacktaxonomy}

This section presents the taxonomy built from 107 unique attack vectors
collected through the review of scientific and grey literature. Following, it
summarizes the results of its validation by domain experts, and the responses of
software developers regarding problem awareness, understandability, and
usefulness of our taxonomy. 

\subsection{Taxonomy of Attacks on \ac{OSS} Supply Chains}\label{attackvectorslist}

The attackers' high-level goal is to conduct a supply chain attack by injecting
malicious code\footnote{This does not only cover the addition of program code
but malicious changes in general, e.g., the introduction of new malicious
dependencies or the (re)introduction of vulnerabilities, e.g., the removal of
authorization checks.} into an \ac{OSS} project such that it is downloaded by
downstream consumers, and executed upon installation or at runtime.
They can target any kind of project (e.g., libraries or word processors), direct
or indirect downstream consumers, as many as possible, or very specific ones.
The latter is possible by conditioning the execution of malicious code, e.g., on
the lifecycle phase (install, test, etc.), application state, operating system,
or properties of the downstream component it has been integrated
into~\cite{ohm2020backstabbers}.

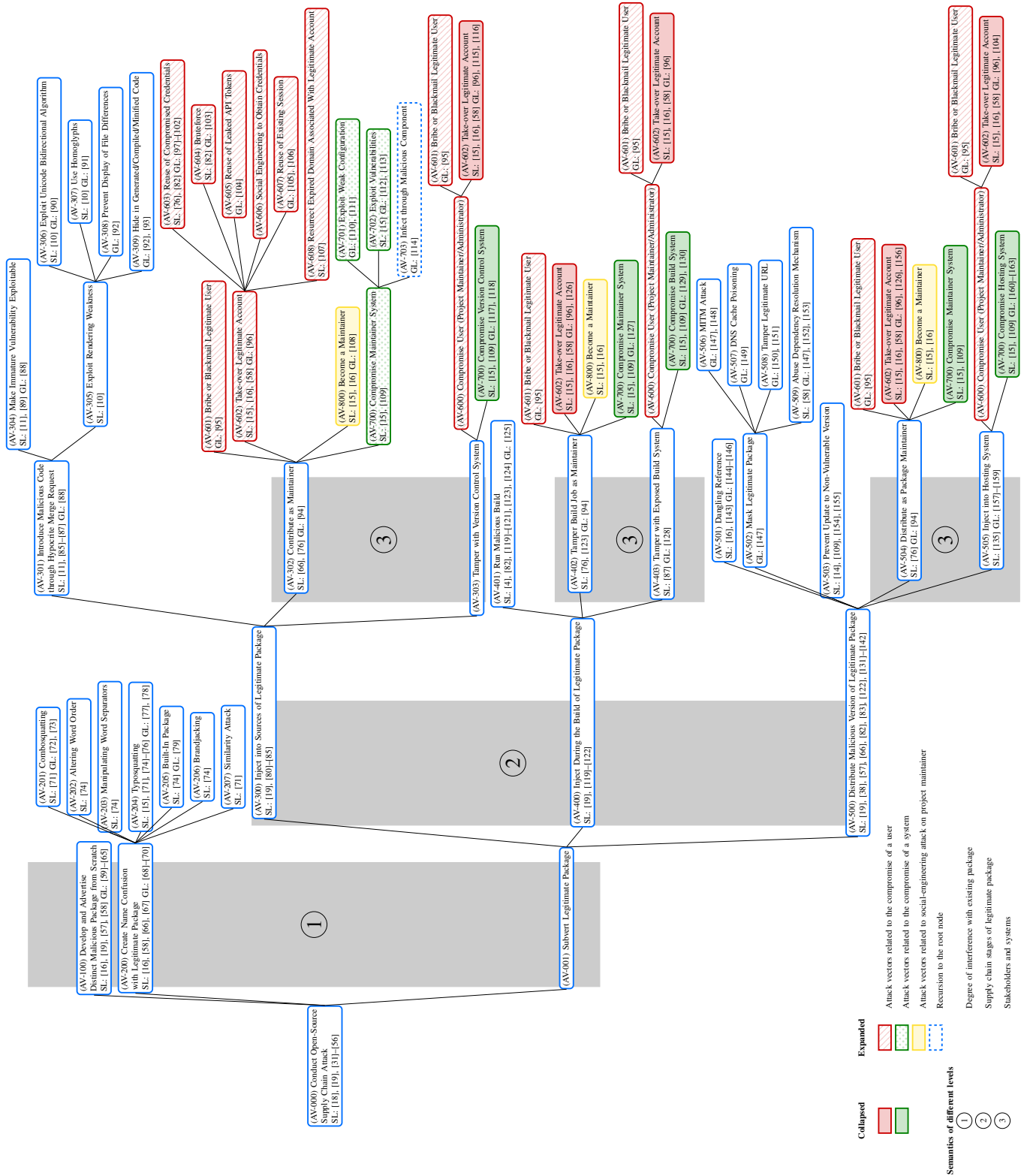
\begin{figure*}[htp]
\begin{adjustbox}{scale=0.44,angle=90}

\sbox0{%
\begin{forest}
    /tikz/every node/.append style={font=\normalsize},
    for tree={%
      treenode,
    parent anchor=east,child anchor=west,
      grow'=east}
    [ (AV-000) Conduct Open-Source \\ Supply Chain Attack \\ SL:\cite{ohm2020backstabbers,10.1145/3459960.3459973,10.1145/3460120.3484736,abdalkareem2020impact,ohmbuildwatch,8909818,8807537,8747065,staicu2018synode,8530068,8587375,10.1145/3239235.3268920,vasilakis2018breakapp,10.1145/3475716.3475769,assal2018security,8234461,10.1145/3098954.3120928,7180277,zhang2015assessing,7332492,7345401,10.1145/2663716.2663755,du2013towards,silic2013information,6107848,5718996,5929123,5428501}
[ (AV-100) Develop and Advertise \\Distinct Malicious Package from Scratch \\ SL: \cite{9402108,236368,sejfia2022practical,ohm2020backstabbers} GL: \cite{trojhorseex1,trojhorseex2,trojhorseex4,trojhorseex3,sharma,thevergeOpenSource,thehackernewsPopularPackage}]
[ (AV-200) Create Name Confusion \\with Legitimate Package \\ SL:\cite{kaplan2021survey,taylor2020spellbound,236368,sejfia2022practical} GL:\cite{sophosPyPIPython,npmjsBlogArchive,govSkcsirtsa20170909pypi}
[ (AV-201) Combosquatting \\ SL:\cite{duclyvu} GL:\cite{mediumDiscordToken,thehackernewsMaliciousLibraries}]
[ (AV-202) Altering Word Order \\ SL:\cite{DBLP:journals/corr/abs-2102-06301}]
[ (AV-203) Manipulating Word Separators \\ SL:\cite{DBLP:journals/corr/abs-2102-06301}]
[ (AV-204) Typosquatting \\ SL:\cite{tschacher2016typosquatting,DBLP:journals/corr/abs-2102-06301,duclyvu,duan2020towards,vaidya2019security} GL:\cite{mediumCryptocurrencyClipboard,snykMaliciousPackages}]
[ (AV-205) Built-In Package \\ SL:\cite{DBLP:journals/corr/abs-2102-06301} GL:\cite{githubApplicationSecurity}] 
[ (AV-206) Brandjacking \\ SL:\cite{DBLP:journals/corr/abs-2102-06301} ]
[ (AV-207) Similarity Attack \\ SL:\cite{duclyvu}]
]
[ (AV-001) Subvert Legitimate Package
[ (AV-300) Inject into Sources of Legitimate Package \\ SL:\cite{ohm2020backstabbers,9402087,la2021windows,8712374,236322, 10.1145/3319535.3345656, 197135}
[ (AV-301) Introduce Malicious Code \\ through Hypocrite Merge Request \\ SL:\cite{wu2021feasibility,goyal2018identifying,197135,7169443} GL:\cite{ReportUniversity}
[ (AV-304) Make Immature Vulnerability Exploitable \\ SL:\cite{wu2021feasibility, 10.1145/2991079.2991103} GL:\cite{ReportUniversity}]
[ (AV-305) Exploit Rendering Weakness \\ SL:\cite{boucher2021trojan} 
[ (AV-306) Exploit Unicode Bidirectional Algorithm \\ SL:\cite{boucher2021trojan} GL:\cite{mitreCVE202142574}]
[ (AV-307) Use Homoglyphs \\ SL:\cite{boucher2021trojan} GL:\cite{mitreCVE202142694}]
[ (AV-308) Prevent Display of File Differences \\ GL:\cite{githubGitHubMortensonprsneaking}]
[ (AV-309) Hide in Generated/Compiled/Minified Code \\ GL:\cite{snykLockfilesSecurity, githubGitHubMortensonprsneaking}]
]
]
[ (AV-302) Contribute as Maintainer \\SL:\cite{kaplan2021survey, vaidya2019security} GL:\cite{garroodMaliciousCode}
[ (AV-601) Bribe or Blackmail Legitimate User \\ GL:\cite{mitreTrustedRelationship},for tree={draw=bostonuniversityred}, for tree={draw=bostonuniversityred, pattern=north east lines, pattern color=bostonuniversityred!20},]
[ (AV-602) Take-over Legitimate Account \\SL:\cite{duan2020towards,236368,sejfia2022practical} GL:\cite{mitreValidAccounts},for tree={draw=bostonuniversityred, pattern=north east lines, pattern color=bostonuniversityred!20},
[ (AV-603) Reuse of Compromised Credentials \\SL:\cite{8712374,vaidya2019security} GL:\cite{mitreUnsecuredCredentials,phpPhpinternalsChanges,portswiggerBackdoorPlanted,snykMaliciousRemote,withatwistStrongpasswordV007,githubCVE201915224Version}]
[ (AV-604) Bruteforce \\SL:\cite{8712374} GL:\cite{mitreBruteForce}]
[ (AV-605) Reuse of Leaked API Tokens\\ GL:\cite{GainedCommit}]
[ (AV-606) Social Engineering to Obtain Credentials]
[ (AV-607) Reuse of Existing Session \\ GL:\cite{certCERTCCVulnerability,mitreCAPECCAPEC60}]
[ (AV-608) Resurrect Expired Domain Associated With Legitimate Account \\SL:\cite{zahan2021weak}]
]
[ (AV-800) Become a Maintainer\\ SL:\cite{duan2020towards,236368} GL:\cite{mediumCompromisedPackage}, top color=bananayellow!20,bottom color=bananayellow!20,draw=bananayellow]
[ (AV-700) Compromise Maintainer System \\ SL:\cite{duan2020towards,203201}, for tree={draw=ao(english), pattern=crosshatch dots, pattern color=ao(english)!20}, 
[ (AV-701) Exploit Weak Configuration\\ GL:\cite{owaspOWASP2017,mitreCWE16Configuration}]
[ (AV-702) Exploit Vulnerabilities \\ SL:\cite{duan2020towards} GL:\cite{mitreExploitPublicFacing,mcclure2010protecting}]
[ (AV-703) Infect through Malicious Component \\ GL:\cite{securelistOperationShadowHammer}, draw=brandeisblue, dashed, fill=white]
]
]
[ (AV-303) Tamper with Version Control System
[ (AV-600) Compromise User (Project Maintainer/Administrator), for tree={draw=bostonuniversityred, pattern=north east lines, pattern color=bostonuniversityred!20},
[ (AV-601) Bribe or Blackmail Legitimate User \\ GL:\cite{mitreTrustedRelationship}]
[ (AV-602) Take-over Legitimate Account \\SL:\cite{duan2020towards,236368,sejfia2022practical} GL:\cite{mitreValidAccounts,tamperingvcsex3,tamperingvcsex2}, for tree={draw=bostonuniversityred, top color=bostonuniversityred!20,, bottom color=bostonuniversityred!20}
]
]
[ (AV-700) Compromise Version Control System \\ SL:\cite{duan2020towards,203201} GL:\cite{zdnetOpensourceProFTPD,tamperingvcsex4},for tree={draw=ao(english), top color=ao(english)!20, bottom color=ao(english)!20},
]
]
]
[ (AV-400) Inject During the Build of Legitimate Package \\ SL:\cite{ohm2020backstabbers,karger2002thirty,wheeler2005countering,9240695,10.1145/3372297.3420015}
[ (AV-401) Run Malicious Build \\ SL:\cite{karger2002thirty,wheeler2005countering,10.1145/2491055.2491070,9240695,8712374,goerigk1999trojan,10.1145/358198.358210} GL:\cite{munoz2020octopus}]
[ (AV-402) Tamper Build Job as Maintainer \\ SL:\cite{10.1145/2491055.2491070,vaidya2019security} GL:\cite{garroodMaliciousCode}
[ (AV-601) Bribe or Blackmail Legitimate User \\ GL:\cite{mitreTrustedRelationship},for tree={draw=bostonuniversityred, pattern=north east lines, pattern color=bostonuniversityred!20},]
[ (AV-602) Take-over Legitimate Account \\SL:\cite{duan2020towards,236368,sejfia2022practical} GL:\cite{mitreValidAccounts,smith2011security},for tree={draw=bostonuniversityred, top color=bostonuniversityred!20,, bottom color=bostonuniversityred!20},
]
[ (AV-800) Become a Maintainer \\ SL:\cite{duan2020towards,236368}, top color=bananayellow!20,bottom color=bananayellow!20,draw=bananayellow]
[ (AV-700) Compromise Maintainer System \\ SL:\cite{duan2020towards,203201} GL:\cite{rossignol2015you}, for tree={draw=ao(english), top color=ao(english)!20, bottom color=ao(english)!20},
]
]
[ (AV-403) Tamper with Exposed Build System \\ SL:\cite{7169443} GL:\cite{arstechnicaAdobeRevoke}
[ (AV-600) Compromise User (Project Maintainer/Administrator),for tree={draw=bostonuniversityred, pattern=north east lines, pattern color=bostonuniversityred!20},
[ (AV-601) Bribe or Blackmail Legitimate User \\ GL:\cite{mitreTrustedRelationship}]
[ (AV-602) Take-over Legitimate Account \\SL:\cite{duan2020towards,236368,sejfia2022practical} GL:\cite{mitreValidAccounts}, for tree={draw=bostonuniversityred, top color=bostonuniversityred!20,, bottom color=bostonuniversityred!20},
]
]
[ (AV-700) Compromise Build System \\SL:\cite{duan2020towards,203201} GL:\cite{webminexploit,paloaltonetworksSiloscapeFirst}, for tree={draw=ao(english), top color=ao(english)!20, bottom color=ao(english)!20},
]
]
]
[ (AV-500) Distribute Malicious Version of Legitimate Package \\ SL:\cite{ohm2020backstabbers,5402538,10.1145/3468264.3468592,vu95py2src,9402108,kaplan2021survey,10.1145/3372297.3420015,Garrett2019DetectingSP,8712374,236322,staicu2018synode,10.5555/2930611.2930648,10.1145/2664243.2664288,tellnes2013dependencies,alhamed2013security,6470829,bellissimo2006secure,10.1145/1137627.1137636,1203227}
[ (AV-501) Dangling Reference \\ SL:\cite{236368,10.1145/3239235.3240501} GL:\cite{danglingreferenceex1,danglingreferenceex2,danglingreferenceex3}]
[ (AV-502) Mask Legitimate Package \\ GL:\cite{almubayed2019practical}
[ (AV-506) MITM Attack \\ GL:\cite{almubayed2019practical,maxMaxComputerTake}]
[ (AV-507) DNS Cache Poisoning \\ GL:\cite{mitreCAPECCAPEC142}]
[ (AV-508) Tamper Legitimate URL \\ GL:\cite{microsoftAttackInception,mitreCWE601Redirection}]
[ (AV-509) Abuse Dependency Resolution Mechanism \\ SL:\cite{sejfia2022practical} GL:\cite{almubayed2019practical,dependencyconfusion,autsoftConfusingDependency}]
]
[ (AV-503) Prevent Update to Non-Vulnerable Version \\ SL:\cite{packagemanagementsecurity,10.1145/1455770.1455841,203201,10.1145/1866307.1866315}]
[ (AV-504) Distribute as Package Maintainer \\ SL:\cite{vaidya2019security} GL:\cite{garroodMaliciousCode}
[ (AV-601) Bribe or Blackmail Legitimate User \\ GL:\cite{mitreTrustedRelationship},for tree={draw=bostonuniversityred, pattern=north east lines, pattern color=bostonuniversityred!20}]
[ (AV-602) Take-over Legitimate Account \\SL:\cite{duan2020towards,236368,sejfia2022practical} GL:\cite{mitreValidAccounts,eslint,smith2011security},for tree={draw=bostonuniversityred, top color=bostonuniversityred!20,, bottom color=bostonuniversityred!20},
]
[ (AV-800) Become a Maintainer \\ SL:\cite{duan2020towards,236368}, top color=bananayellow!20,bottom color=bananayellow!20,draw=bananayellow]
[ (AV-700) Compromise Maintainer System \\ SL:\cite{duan2020towards,203201}, for tree={draw=ao(english), top color=ao(english)!20, bottom color=ao(english)!20},
]
]
[ (AV-505) Inject into Hosting System \\ SL:\cite{10.5555/2930611.2930648} GL:\cite{broadcomEndpointProtection,linuxmintBewareHacked,krebsonsecurityBuryMajor}
[ (AV-600) Compromise User (Project Maintainer/Administrator),for tree={draw=bostonuniversityred, pattern=north east lines, pattern color=bostonuniversityred!20},
[ (AV-601) Bribe or Blackmail Legitimate User \\ GL:\cite{mitreTrustedRelationship}]
[ (AV-602) Take-over Legitimate Account \\SL:\cite{duan2020towards,236368,sejfia2022practical} GL:\cite{mitreValidAccounts,GainedCommit}, for tree={draw=bostonuniversityred, top color=bostonuniversityred!20,, bottom color=bostonuniversityred!20}
]
]
[ (AV-700) Compromise Hosting System \\ SL:\cite{duan2020towards,203201} GL:\cite{phpcomposer,packagistrce,rubygemsrce,welivesecuritySupplychainAttack}, for tree={draw=ao(english), top color=ao(english)!20, bottom color=ao(english)!20},
]
]
]
]
]
    \end{forest}}%

\sbox1{
  
\begin{tabular}{ccl}
    
    \textbf{Collapsed} & \textbf{Expanded} & \\
    \\
    \tikz{\draw[bostonuniversityred,ultra thick, fill=bostonuniversityred!20] (0,0) rectangle (1,0.5)} &\tikz{\draw[bostonuniversityred,pattern=north east lines, pattern color=bostonuniversityred!20,ultra thick] (0,0) rectangle (1,0.5)} & Attack vectors related to the compromise of a user\\
    \tikz{\draw[ao(english),fill=ao(english)!20,ultra thick] (0,0) rectangle (1,0.5)} &\tikz{\draw[ao(english),pattern=crosshatch dots, pattern color=ao(english)!20,ultra thick] (0,0) rectangle (1,0.5)} & Attack vectors related to the compromise of a system\\
     &\tikz{\draw[bananayellow,fill=bananayellow!20,ultra thick] (0,0) rectangle (1,0.5)} & Attack vectors related to social-engineering attack on project maintainer\\
     &\tikz{\draw[brandeisblue,dashed,ultra thick] (0,0) rectangle (1,0.5)} & Recursion to the root node\\
     \textbf{Semantics of different levels} & &\\
     \tikz{\node[draw,circle] {1};} & & Degree of interference with existing package
    \\
    \tikz{\node[draw,circle] {2};} & & Supply chain stages of legitimate package
    \\
    \tikz{\node[draw,circle] {3};} & & Stakeholders and systems
    \\
    
    \end{tabular}

    }%
    \sbox2{
        
        \tikz{\filldraw[black!20] (0,10) rectangle (5,33)}

    }

    \sbox3{
        \tikz{\filldraw[black!20] (0,0) rectangle (5,24)}
    }
    \sbox4{
      
        \tikz{\filldraw[black!20] (0,0) rectangle (5,6)}
    }

    \sbox5{
        \tikz{\filldraw[black!20] (0,0) rectangle (5,6)}
    }
    \sbox6{
        
        \tikz{\filldraw[black!20] (0,0) rectangle (5,8.6)}
    }

    \sbox7{
        \tikz{\node[draw,circle] {\Huge{3}};}
    }
    \sbox8{
        \tikz{\node[draw,circle] {\Huge{2}};}
    }
    \sbox9{
        \tikz{\node[draw,circle] {\Huge{1}};}
    }

\begin{tikzpicture}
  \node (forest){\usebox0};
  \node[] at (-16,-17) {\usebox1};
  \begin{scope}[on background layer]
    \node[rounded corners] at (-14.5, 8) {\usebox2 } ;
    \node[rounded corners] at (-14.5, 8) {\usebox9 } ;
    \node[rounded corners] at (-8, -1.5) {\usebox3 } ;
    \node[rounded corners] at (-8, 0) {\usebox8 } ;
    \node[rounded corners] at (1, -4.7) {\usebox4 } ;
    \node[rounded corners] at (1, -4.7) {\usebox7 } ;
    \node[rounded corners] at (1, -17.4) {\usebox5 } ;
    \node[rounded corners] at (1, -17.4) {\usebox7 } ;
    \node[rounded corners] at (1, 5.4) {\usebox6 } ;
    \node[rounded corners] at (1, 5.2) {\usebox7 } ;
\end{scope}
  
\end{tikzpicture}

\end{adjustbox}

    \caption{Refined version of the taxonomy for \ac{OSS} supply chain attacks.
    It takes the form of an attack tree with the attacker's top-level goal to
    inject malicious code into open-source project artifacts consumed and
    executed by downstream users. This version reflects the feedback of 17
    domain experts on the initial version, collected through an online survey.
    Subtrees for user and system compromises exist multiple times, only their
    first occurrence is expanded. The grey, numbered rectangles illustrate the
    different criteria used for structuring the tree. Each node has references to 
    both Scientific Literature (SL) and Gray Literature (GL).}
    \label{fig:attacktreetaxonomy}

\end{figure*}

The entire taxonomy unfolding below this high-level goal is depicted in
Figure~\ref{fig:attacktreetaxonomy} and summarized hereafter, whereby the
1st-level child nodes of the tree reflect different degrees of interference
with existing packages.



\textbf{Develop and Advertise Distinct Malicious Package from Scratch}
covers the creation of a new \ac{OSS} project, with the intention
to use it for spreading malicious code from the beginning or at a later point
in time. Besides creating the project, the attacker is required to advertise the
project to attract victims. Real-world examples affect PyPI, npm, Docker Hub or
NuGet~\cite{ohm2020backstabbers,trojhorseex1,trojhorseex2,trojhorseex4,trojhorseex3,
sharma,thevergeOpenSource,thehackernewsPopularPackage}.




\textbf{Create Name Confusion with Legitimate Package}
covers attacks that consist of creating project or artifact names that resemble
legitimate ones, suggest trustworthy authors, or play with common naming
patterns. Once a suitable name is found, the malicious artifact is deployed,
e.g., in a source or package repository, in the hope of being consumed by 
downstream users. As the deployment does not interfere with the resources of the
project that inspired the name (e.g., legitimate code repository, maintainer
accounts) the attack is relatively cheap.

Child nodes of this attack vector relate to sub-techniques applying different
modifications to the legitimate project name:
\emph{Combosquatting}~\cite{DBLP:journals/corr/abs-2102-06301} adds pre or
post-fixes, e.g., to indicate project maturity (\texttt{dev} or
\texttt{rc}) or platform compatibility (\texttt{i386}).
\emph{Altering Word Order}~\cite{DBLP:journals/corr/abs-2102-06301} re-arranges
the word order (\texttt{test-vision-client} vs.
\texttt{client-vision-test}).
\emph{Manipulating Word Separators}~\cite{DBLP:journals/corr/abs-2102-06301}
alters or adds word separators like hyphens (\texttt{setup-tools} vs.
\texttt{setuptools}).
\emph{Typosquatting}~\cite{DBLP:journals/corr/abs-2102-06301,tschacher2016typosquatting,ohm2020backstabbers,herr2021breaking,duan2020towards,duclyvu}
exploits typographical errors (\texttt{dajngo} vs. \texttt{django}).
\emph{Built-In Package}~\cite{DBLP:journals/corr/abs-2102-06301} replicates
well-known names from other contexts, e.g., built-in packages or modules of a
programming language (\texttt{subprocess} for Python).
%
\emph{Brandjacking}~\cite{brandjacking} creates the impression a
package comes from a trustworthy author (\texttt{twilio-npm}).
\emph{Similarity Attack}~\cite{iqttyposquatting} creates a misleading name in a way different from the
previous categories (\texttt{request} vs. \texttt{requests}).

\textbf{Subvert Legitimate Package} covers all attacks aiming to corrupt an
existing, legitimate project, which requires compromising one or more of its
numerous resources depicted in Figure~\ref{fig:sdlc}. As a result, this subtree
is much larger compared to the previous ones, esp. because subtrees related to
user and system compromises occur multiple times in the different supply chain
stages. The remainder of this section is dedicated to sub-techniques of this
first-level node.

\subsubsection*{Inject into Sources of Legitimate Package}
\label{injectintosources}
It relates to the injection of malicious code into a project's
codebase. For the attacker, this has the advantage to affect all downstream
users, no matter whether they consume sources or pre-built binary artifacts (as
part of the codebase, the malicious code will be included during project builds
and binary artifact distribution).


This vector has several sub-techniques. Taking the role of contributors,
attackers can use \emph{hypocrite merge requests} to turn immature
vulnerabilities into exploitable ones~\cite{wu2021feasibility}, or exploit IDE
rendering weaknesses to hide malicious code, e.g., through the use of Unicode
homoglyphs and control characters~\cite{boucher2021trojan}, or the hiding and
suppression of code
differences~\cite{prsneaking}.
%
To \emph{contribute as maintainer} requires to obtain the privileges necessary
for altering the legitimate project's codebase, which can be achieved in
different ways. Using \ac{SE} techniques on legitimate project maintainers
~\cite{giovanini2021leveraging, eventstream}, 
by \emph{taking over legitimate accounts} (e.g., reusing compromised
credentials~\cite{accountakeover}), or by \emph{compromising the maintainer
system} (e.g., exploiting vulnerabilities~\cite{mcclure2010protecting}). The
latter can also be achieved through a malicious (\ac{OSS}) component, e.g., IDE
plugin, which is reflected through a recursive reference to the root node.

The legitimate project's codebase can also be altered by \emph{tampering with
its \ac{VCS}}, thus, bypassing a project's established contribution workflows. For
instance, by compromising system user
accounts~\cite{tamperingvcsex2,tamperingvcsex3}, or by exploiting
configuration/software
vulnerabilities~\cite{tamperingvcsex1,tamperingvcsex4,phpsrchack}, an attacker could
access the codebase in insecure ways.

\subsubsection*{Inject During the Build of Legitimate Package}
\label{injectduringbuild}
Greatly facilitated by language-specific package managers like Maven or Gradle
for Java, it became common to download pre-built components from package
repositories rather than \ac{OSS} project's source code from its \ac{VCS}.
Therefore, the injection of malicious code can happen during the build of such
components before their
publication~\cite{chess2007attacking,karger2002thirty,wheeler2005countering}.
Though the spread is limited compared to injecting into sources, the advantage
for the attacker is that the detection of malicious code inside pre-built
packages is typically more difficult, especially for compiled programming
languages.
One sub-technique is \emph{running a malicious build job} to tamper with system
resources shared between build jobs of multiple
projects~\cite{10.1145/2491055.2491070} (e.g., the infection of Java archives in
NetBeans projects~\cite{munoz2020octopus}).
An attacker can also \emph{tamper the build job as maintainer}, e.g., by taking
over legitimate maintainer accounts, becoming a maintainer, or compromising their
systems (cf. XCodeGhost malware~\cite{rossignol2015you}). Similarly, the
attacker could comprise build systems, esp. online accessible ones, e.g., by
compromising administrator accounts~\cite{smith2011security} or exploiting
vulnerabilities~\cite{teslajenkins,webminexploit}.

\subsubsection*{Distribute Malicious Version of Legitimate Package}
Pre-built components are often hosted on well-known package repositories like
PyPI or npm, but also on less popular repositories with a narrower scope. In
addition, the components can be mirrored remotely or locally, made available
through \ac{CDN}s (e.g., in the case of JavaScript libraries), or cached in
proxies. This attack vector and its sub-techniques cover all cases where
attackers tamper with mechanisms and systems involved in the hosting,
distribution, and download of pre-built packages.

\emph{Dangling references} (re)uses resource identifiers of orphaned
projects~\cite{danglingreferenceex1,danglingreferenceex2,danglingreferenceex3},
e.g., names or URLs.
\emph{Mask legitimate package}~\cite{almubayed2019practical} targets package
name or URL resolution mechanisms and download connections. Their goal is the
download of malicious packages by compromising resources external to the
legitimate project. This includes \ac{MITM} attacks, DNS cache poisoning, or
tampering with legitimate URLs directly at the
client~\cite{microsoftmobileexample}. Particularly, package managers follow a
(configurable) resolution strategy to decide which package version to download,
from where, and the order of precedence when contacting multiple repositories.
Attackers can \emph{abuse such resolution mechanisms} and their
configurations~\cite{dependencyconfusion,aconfusingdependency}.
Attackers can also \emph{prevent updates to non-vulnerable versions} by
manipulating package metadata~\cite{10.1145/1455770.1455841}, e.g., by indicating
an unsatisfiable dependency for newer versions of a legitimate package.
Finally, the involvement of systems and users in package distribution results in
attack vectors similar to previous ones. Attackers can take the role of
legitimate maintainers, thus, \textit{distribute as maintainer}, e.g., by taking
over package maintainer accounts (e.g., \texttt{eslint}~\cite{eslint}), the
second most common attack vector after typosquatting~\cite{ohm2020backstabbers}.
They can also compromise maintainer systems, or directly \emph{inject into the
hosting system}, e.g., by compromising administrator accounts~\cite{codecov} or
exploiting vulnerabilities~\cite{phpcomposer,npmvuln2021,rubygemsrce,packagistrce}.

\begin{tcolorbox}
    \textbf{Response to RQ1.1}: Through the review of 183 scientific papers as
    well as grey literature, we identified and generalized 107 unique
    attack vectors on \ac{OSS} supply chains, supported by 94 real-world
    attacks or vulnerabilities.
    \end{tcolorbox}

\subsection{Validation and Assessment by Domain Experts}\label{sec:expertsurvey}


The initial version of the taxonomy was
validated and assessed by \textbf{17 domain experts}. Their feedback has been retrofitted
resulting in the taxonomy depicted in
Figure~\ref{fig:attacktreetaxonomy}.

\subsubsection*{Validation} This section reports expert feedback on the comprehensiveness of attack vectors,
and the correctness, comprehensibility, and usefulness of the taxonomy.
    
    
Before having seen the taxonomy in its entirety, the tree-testing
required experts to assign attack vectors to the first level nodes of the
initial taxonomy.
Over a total of 311 assignments by all experts, 234 (75\%) matched the structure
of the initial taxonomy, while 77 (25\%) did not, which shows an overall
agreement on the structure. 


Following, the experts were presented with the initial version of the entire
taxonomy, and asked to assess different qualities using a Likert scale ranging
from 1 (low) to 5 (high).

14 (82\%) experts agreed to the overall structure with a rating of 4 or 5,
slightly higher compared to the results of the tree-test. \mm{which is that result? cross-ref?}
This could be
due to some node names not being self-explanatory enough when shown with too
little context.

Experts were further asked to rate the correctness of the taxonomy's 1st-level
nodes in regards to naming, tree location, and sub-tree structure. 
All the first-level nodes received an overall good agreement with naming, categorization, and sub-tree
structure, except \textit{Develop and Advertise Distinct Malicious Package from Scratch}. The latter only
received neutral feedback on its sub-tree, a light agreement with its
categorization, and a clear disagreement with its initial naming. 


12 (71\%) of the experts agreed with the \textbf{completeness of the attack tree}.
\mm{This sentence is quite isolated. As we do not have info about what is missing and the percentage of  expert saying is not complete is quite high (30\%) it would simple remove the sentence. }

\subsubsection*{Usefulness and Use-Cases}
    
In this part of the questionnaire, experts rated the usefulness and possible use
cases of the proposed taxonomy.

15 (88\%) rated the usefulness of the taxonomy to \textit{understand the
attack surface of the \ac{OSS} supply chain} with a 4 or 5. 
Fewer experts considered it being useful to understand \textit{attacker tactics and techniques} (12 (71\%)) or
\textit{attackers' cost/benefits considerations} (5 (29\%)).
    

Regarding the expert options about possible uses of the proposed taxonomy,
the Top-3 use-cases are \textit{threat modeling},
\textit{awareness and training} and \textit{risk assessment}. Another possible
use-case, though not included in the survey, is to scope penetration tests.
\mm{if space is need, you can remove the figure and put the percentage here: see that the figure shows only 4 uses and the text 3.}
    

\subsection{Validation and Assessment by Developers}

The initial version of the taxonomy has also been validated and assessed by
\textbf{134 software developers} in regards to the awareness of main attack
vectors (1st-level taxonomy nodes), whether those are mitigated (by themselves
or their organization), and \textendash~optionally \textendash~the understandability and utility of
the taxonomy. 

\subsubsection*{Awareness about Attack Vectors}


The awareness of main attack vectors ranged from 120 (90\%) for
\textit{Develop and Advertise Distinct Malicious Package from Scratch} to 86 (64\%) for \textit{Inject
During the Build of Legitimate Package}.


For all but one vector, the majority of respondents answered not to know whether
they are protected. Only for \textit{Develop and Advertise Distinct Malicious Package from Scratch}, the
majority believes in being protected (52\%). For both vectors \textit{Inject
During the Build} and \textit{Distribute Malicious Version}, 19 (14\%)
respondents were sure that no protection exists.

\subsubsection*{Taxonomy Understandability and Utility Assessment}

Among the 134 participants, 53 (40\%) decided to perform the optional
assessment of the taxonomy's understandability and utility to understand the
supply chain's attack surface.
Considering a rating of 4 or 5, 41 (77\%) found the taxonomy understandable
and 46 (87\%) recognized it as a useful means to create awareness.

\begin{tcolorbox}
\textbf{Response to RQ1.2}: The proposed taxonomy of attacks on \ac{OSS} supply
chains takes the form of an attack tree covering all 107 vectors identified
beforehand.

Its validation by 17 domain experts and 134 software developers showed overall
agreement with structure and naming, comprehensiveness, comprehensibility, and
suitability for use-cases like threat modeling, awareness, training, or risk
assessment.



\end{tcolorbox}

\section{Safeguards and their Assessments}
\label{section:safeguards}
Subsection~\ref{section:safeguardslist} starts with a short overview about safeguards
against \ac{OSS} supply chain attacks, which were identified through literature
review and generalized to become agnostic of specific prg. languages or
ecosystems. Subsections~\ref{section:expertsurveySafeguards}~and
~\ref{section:developersurveySafeguards} report the results of the two surveys
conducted with domain experts and software developers to validate and assess the
safeguards regarding different qualities, e.g., utility or cost.

\subsection{List of Safeguards}\label{section:safeguardslist} In total, we
identified 33 safeguards that partially or completely mitigate the
before-mentioned attack vectors. Both implementation and use of those safeguards
can incur non-negligible costs, also depending on the specifics of prg.
languages and ecosystems at hand. Thus, the selection, combination and
implementation of safeguards require careful planning and design, to balance
required security levels and costs.

The complete list of safeguards can be found in Table~\ref{tab:safeguards} of
Appendix~\ref{appendix:listofsafeguards}, including a classification after
control type. All safeguards are mapped to the vector(s) they mitigate, some to
the top-level goal due to (partially) addressing all vectors (e.g., establishing
a vetting process), others to more specific subgoals. Some safeguards can be
implemented by one or more stakeholders, while others require the involvement of
multiple ones to be effective (e.g., signature creation and verification).


\textbf{Common Safeguards}
comprises 4 countermeasures that require all stakeholders to become active,
i.e., project maintainers, open-source consumers, and administrators (service
providers). For example, a detailed \ac{SBOM} has to be produced and maintained
by the project maintainer~\cite{deliveruncompromised}, ideally using automated
\ac{SCA} tools. Following, the \ac{SBOM} must be securely hosted and
distributed by package repositories, and carefully checked by downstream users
in regards to their security, quality, and license requirements.

\textbf{Safeguards for Project Maintainers and Administrators}
comprises eight safeguards. \textit{Secure authentication},
for instance, suggests service providers to offer \ac{MFA} or enforce strong
password policies, while project maintainers should follow authentication
best-practices, e.g., use \ac{MFA} where available, avoid password reuse, or
protect sensitive tokens.

\textbf{Safeguards for Project Maintainers}
includes seven countermeasures. Generally, \ac{OSS} projects use
hosted, publicly accessible \ac{VCS}s. Maintainers should then, e.g., conduct
careful \textit{merge request reviews} or enable \textit{branch protection
rules} for sensitive project branches to avoid malicious code
contributions. 
As project builds may still happen on maintainers' workstations, they are
advised to use \textit{dedicated build services}, esp. \textit{ephemeral
environments}~\cite{slsaframework}. Additionally, they may \textit{isolate build
steps}~\cite{10.1145/2491055.2491070} such that they cannot tamper with the
output of other build steps.


\textbf{Safeguards for Administrators and Consumers} comprises
five countermeasures. For example, both package repository administrators
and consumers can opt for \textit{building packages directly from source
code}~\cite{googleopensource}\hp{Henrik's reddit post}, rather than accepting
pre-built artifacts. If implemented by package repositories, this would reduce
the risk of subverted project builds. If implemented by consumers, this would
eliminate all risks related to the compromise of 3rd-party build services and
package repositories, as they are taken out of the picture.

\textbf{Safeguards for Consumers}
includes nine countermeasures that may be employed by the
downstream users. The consumers of \ac{OSS} packages may reduce the impact of
malicious code execution when consuming by \textit{isolating the code
and/or sandboxing} it. Another example is the \textit{establishment of internal
repository mirrors}~\cite{microsoftwhitepaper} of vetted components.

\begin{tcolorbox}
    \textbf{Response to RQ2.1}: We identified 33 general safeguards to be used
    by the different stakeholders, mostly detective or preventive ones, and mapped
    them to the node(s) of the attack tree they mitigate partially or fully.
\end{tcolorbox}

\subsection{Experts Validation and Assessment}
\label{section:expertsurveySafeguards}
This section presents the feedback of 17 experts regarding the safeguards'
utility to mitigate risks, and their associated costs for implementation and
continued use.

In summary, almost all the safeguards received medium to high utility ratings,
while the cost ratings range from low (i.e., minimum mean value of 2.0) to very
high (i.e., maximum mean value of 4.8).

Table~\ref{tab:safeguards2} provides all feedback collected for the 33
safeguards, following a discussion of safeguards with the highest, respectively
lowest \ac{U/C} ratios, and some other interesting cases.

\textbf{High \ac{U/C} ratio.} Both \textit{Protect production branch},
\textit{Remove un-used dependencies} and \textit{Version pinning} show the
highest \ac{U/C} ratio, thus, are considered to be useful and cheap controls.
The use of \textit{Resolution Rules} also shows a good \ac{U/C} ratio, even
though one expert highlighted that "very few projects" use them, and that the
implementation would require the modification of all package managers.
On average, \textit{Preventive Squatting} only received neutral ratings (3.1 for
utility and 2.9 for cost) and also raised some concerns: two of the experts highlighted
that it could be good to "try to prevent name squatting, but hard to fully
enforce" also due to legitimate reasons for similar names (e.g., to help
consumers identify package relationships).

\textbf{Low \ac{U/C} ratio.} \textit{Build Dependencies from Sources},
reportedly used by
Google~\cite{googleopensource},
received a very low utility rating (mean and median of 3.0) and overall the
lowest \ac{U/C} ratio. Considering that its use would prune the subtrees of both
vectors \textit{Inject During the Build} and \textit{Distribute Malicious
Version}, we expected a higher utility rating. One expert claimed that "building
from source only helps if someone scans and reviews the code". Possibly
referring to flaky builds~\cite{flakybuilds}, another expert highlighted that
"rebuilding software from source can sometimes introduce problems".

\textit{Merge Request Reviews} received the highest average utility rating
(4.6), which could be because if malicious code is injected into the sources, it
is guaranteed to arrive at consumers, no matter how they consume it.

\textit{Reproducible Builds} received a very high utility rating (5) from 10
participants (58.8\%), but also a high-cost rating (4 or 5) from 12 (70.6\%).
One expert commented that a "reproducible build like used by Solarwinds now, is
a good measure against tampering with a single build system" and another claimed
this "is going to be the single, biggest barrier".

\textit{Scoped Packages}, proposed as an effective safeguard against
\textit{Abuse of Dependency Resolution
mechanisms}~\cite{dependencyconfusionschibsted,microsoftwhitepaper}, mostly
received neutral ratings (3) for both utility and cost.

\begin{tcolorbox}
    \textbf{Response to RQ2.2}:  We have qualitatively assessed the utility and
    costs of the 33 safeguards by surveying 17 experts. The three safeguards
    \textit{Protect production branch}, \textit{Remove un-used dependencies} and
    \textit{Version pinning} showed the best \ac{U/C} ratio while
    \textit{Build dependencies from sources} showed the worst.
\end{tcolorbox}


\subsection{Developers Validation and Assessment}
\label{section:developersurveySafeguards} 

In this optional part of the survey, developers were asked to assess the usage
and costs of a subset of safeguards that were selected according to the
stakeholders' roles exercised in their daily work (collected in the demographic
part). Among the total of 134 respondents, 30 assessed the \textit{Common
Safeguards}, 5 the \textit{Safeguards for Project Maintainers}, 4 the
\textit{Safeguards for Maintainers and Administrators}, 24 the
\textit{Safeguards for Administrators and Consumers}, and 22 the
\textit{Safeguards for Consumers}. Complete results are shown in Table~\ref{tab:safeguards2}


\textit{Remove un-used dependencies} is frequently used by developers, which
contrasts with the observations of Soto-Valerio et al.~\cite{depclean},
who found that many Java projects had bloated (un-used) dependencies.
Other countermeasures that appear to be widely used among the respondents are
\textit{Version pinning} and \textit{Open-source vulnerability scanners}, the
latter of which does not only address attacks, but also the use of dependencies
with known vulnerabilities.
\hp{OB/MM: Ok to not always use the exact name?}

Concerning the attack vector \textit{Create Name Confusion}, where 70\% of the
developers claimed to be aware of the problem, we can observe that corresponding
safeguards \textit{Typo guard/Typo detection} and \textit{Preventive squatting
the released package} are only used by a minority of respondents.

It is also noteworthy to mention that developers' cost ratings generally coincide with those of the domain expert. Surprising exceptions are \emph{Application Security Testing} and \emph{Enstablish vetting process for Open-Source components hosted in internal/public repositories}, both having a median of 3 from developers, compared to a median of 5 from experts.

\begin{tcolorbox}
    \textbf{Response to RQ2.3}: 134 software developers provided feedback on the
    use of safeguards. The three most-used ones are \textit{Remove un-used
    dependencies}, \textit{Version pinning} and \textit{Integrate Open-Source
    vulnerability scanner into \ac{CI}/\ac{CD} pipeline}.
\end{tcolorbox}

\begin{table*}[!hbtp]

    \centering
    \scalebox{0.94}{
    \begin{tabular}{r|p{6.6cm}lllllllllll|}
    \toprule
    
    \multicolumn{1}{l}{} & \multicolumn{7}{c}{\textbf{Experts}}  & \multicolumn{4}{c}{\textbf{Developers}} \\  \cmidrule(lr){2-8} \cmidrule(lr){9-13}
    
    \multicolumn{1}{l}{} & \multicolumn{3}{c}{\textbf{Utility}}  & \multicolumn{3}{c}{\textbf{Cost}} & \multicolumn{1}{c}{} &  & 
    \multicolumn{3}{c}{\textbf{Cost}}\\  \cmidrule(lr){2-4}  \cmidrule(lr){5-7}  \cmidrule(lr){10-13}

    
    \multicolumn{1}{c}{\multirow{-2}{*}{\textbf{Safeguard }}} & \multicolumn{1}{l}{\rotatebox[origin=c]{60}{\textbf{Mean}}} & \multicolumn{1}{l}{\rotatebox[origin=c]{60}{\textbf{Median}}} & \multicolumn{1}{l}{}  & \multicolumn{1}{l}{\rotatebox[origin=c]{60}{\textbf{Mean}}} & \multicolumn{1}{l}{\rotatebox[origin=c]{60}{\textbf{Median}}} & \multicolumn{1}{l}{} & \multicolumn{1}{c}{\rotatebox[origin=c]{60}{\textbf{Mean U/C }}}  & \multicolumn{1}{c}{\rotatebox[origin=c]{60}{\textbf{Usage}}} & 
    \multicolumn{1}{l}{\rotatebox[origin=c]{60}{\textbf{Mean}}} & \multicolumn{1}{l}{\rotatebox[origin=c]{60}{\textbf{Median}}} & \multicolumn{1}{l}{}\\

    \midrule
    \multicolumn{1}{p{6.6cm}}{Protect production branch~\cite{goyal2018identifying,197135}} &
    \multicolumn{1}{c}{\cellcolor{NextBlue!20}{4.2}} &
    \multicolumn{1}{c}{\cellcolor{NextBlue!20}{4.0}} &
    \multicolumn{1}{c}{\mybarc{0}\mybarc{0}\mybarc{5}\mybarc{4}\mybarc{8}} &
    \multicolumn{1}{c}{\cellcolor{NextBlue}{2.0}}&
    \multicolumn{1}{c}{\cellcolor{NextBlue}{2.0}} &
    \multicolumn{1}{c}{\mybarc{7}\mybarc{6}\mybarc{6}\mybarc{2}\mybarc{0}} &
    \multicolumn{1}{c}{ \textbf{2.10}} &
    \multicolumn{1}{c}{Y\mybarc{3}\mybarc{2}N} &
    \multicolumn{1}{c}{\cellcolor{NextBlue}{1.8}}&
    \multicolumn{1}{c}{\cellcolor{NextBlue}{2.0}} &
    \multicolumn{1}{c}{\mybarc{2}\mybarc{2}\mybarc{1}\mybarc{0}\mybarc{0}}\\ 
    
    \multicolumn{1}{p{6.6cm}}{Remove un-used dependencies~\cite{depclean}} &
    \multicolumn{1}{c}{\cellcolor{NextBlue!20} 4.3} &
    \multicolumn{1}{c}{\cellcolor{NextBlue!20} 5.0} &
    \multicolumn{1}{c}{\mybarc{0}\mybarc{1}\mybarc{3}\mybarc{3}\mybarc{10}} &
    \multicolumn{1}{c}{\cellcolor{NextBlue}{2.1}}&
    \multicolumn{1}{c}{\cellcolor{NextBlue}{2.0}} &
    \multicolumn{1}{c}{\mybarc{6}\mybarc{5}\mybarc{4}\mybarc{2}\mybarc{0}}&
    \multicolumn{1}{c}{2.05}& \multicolumn{1}{c}{Y\mybarc{21}\mybarc{2}N} &
    \multicolumn{1}{c}{\cellcolor{NextBlue}{2.0}}&
    \multicolumn{1}{c}{\cellcolor{NextBlue}{2.0}} &
    \multicolumn{1}{c}{\mybarc{7}\mybarc{9}\mybarc{5}\mybarc{0}\mybarc{1}}\\ 
    
    \multicolumn{1}{p{6.6cm}}{Version
    pinning~\cite{dependencyconfusionschibsted,duan2020towards,microsoftwhitepaper}}&
    \multicolumn{1}{c}{\cellcolor{NextBlue!20}{3.7}} &
    \multicolumn{1}{c}{\cellcolor{NextBlue!60}{3.0}} &
    \multicolumn{1}{c}{\mybarc{0}\mybarc{2}\mybarc{7}\mybarc{2}\mybarc{6}} &
    \multicolumn{1}{c}{\cellcolor{NextBlue}{2.2}}&
    \multicolumn{1}{c}{\cellcolor{NextBlue}{2.0}} &
    \multicolumn{1}{c}{\mybarc{8}\mybarc{3}\mybarc{2}\mybarc{3}\mybarc{1}} &
    \multicolumn{1}{c}{1.68} & \multicolumn{1}{c}{Y\mybarc{20}\mybarc{2}N} &
    \multicolumn{1}{c}{\cellcolor{NextBlue}{2.1}}&
    \multicolumn{1}{c}{\cellcolor{NextBlue}{2.0}} &
    \multicolumn{1}{c}{\mybarc{6}\mybarc{10}\mybarc{4}\mybarc{2}\mybarc{0}}\\ 
    
    \multicolumn{1}{p{6.6cm}}{Dependency resolution rules} &
    \multicolumn{1}{c}{\cellcolor{NextBlue!20}{4.1}} &
    \multicolumn{1}{c}{\cellcolor{NextBlue!20}{4.0}} &
    \multicolumn{1}{c}{\mybarc{0}\mybarc{1}\mybarc{4}\mybarc{5}\mybarc{7}} &
    \multicolumn{1}{c}{\cellcolor{NextBlue!60}{2.6}} &
    \multicolumn{1}{c}{\cellcolor{NextBlue!60}{3.0}} &
    \multicolumn{1}{c}{\mybarc{2}\mybarc{6}\mybarc{6}\mybarc{2}\mybarc{1}} &
    \multicolumn{1}{c}{1.58}& \multicolumn{1}{c}{Y\mybarc{9}\mybarc{13}N} &
    \multicolumn{1}{c}{\cellcolor{NextBlue!60}{2.7}}
    &  \multicolumn{1}{c}{\cellcolor{NextBlue!60}{3.0}} &
    \multicolumn{1}{c}{\mybarc{3}\mybarc{7}\mybarc{6}\mybarc{4}\mybarc{2}}\\ 
    
    \multicolumn{1}{p{6.6cm}}{User account management~\cite{10.5555/2930611.2930648}} &
    \multicolumn{1}{c}{\cellcolor{NextBlue!20}{3.9}} &
    \multicolumn{1}{c}{\cellcolor{NextBlue!20}{4.0}} &
    \multicolumn{1}{c}{\mybarc{0}\mybarc{0}\mybarc{6}\mybarc{7}\mybarc{4}} &
    \multicolumn{1}{c}{\cellcolor{NextBlue!60}{2.6}} &
    \multicolumn{1}{c}{\cellcolor{NextBlue!60}{3.0}} &
    \multicolumn{1}{c}{\mybarc{4}\mybarc{3}\mybarc{7}\mybarc{1}\mybarc{2}} &
    \multicolumn{1}{c}{ 1.50} & \multicolumn{1}{c}{Y\mybarc{3}\mybarc{1}N} &
    \multicolumn{1}{c}{\cellcolor{NextBlue}{2.3}}&
    \multicolumn{1}{c}{\cellcolor{NextBlue}{2.5}} &
    \multicolumn{1}{c}{\mybarc{1}\mybarc{1}\mybarc{2}\mybarc{0}\mybarc{0}}\\ 
    
    \multicolumn{1}{p{6.6cm}}{Secure authentication (e.g., \ac{MFA}, password
    recycle, session timeout, token protection)~\cite{duan2020towards,kaplan2021survey}} &
    \multicolumn{1}{c}{\cellcolor{NextBlue!20}{4.3}}  &
    \multicolumn{1}{c}{\cellcolor{NextBlue!20}{5.0}} &
    \multicolumn{1}{c}{\mybarc{0}\mybarc{1}\mybarc{2}\mybarc{4}\mybarc{10}} &
    \multicolumn{1}{c}{\cellcolor{NextBlue!60}{2.9}} &
    \multicolumn{1}{c}{\cellcolor{NextBlue!60}{3.0}} &
    \multicolumn{1}{c}{\mybarc{1}\mybarc{5}\mybarc{6}\mybarc{5}\mybarc{0}}&
    \multicolumn{1}{c}{ 1.48} & \multicolumn{1}{c}{Y\mybarc{4}\mybarc{0}N} &
    \multicolumn{1}{c}{\cellcolor{NextBlue}{2.5}}&
    \multicolumn{1}{c}{\cellcolor{NextBlue!60}{3.0}} &
    \multicolumn{1}{c}{\mybarc{1}\mybarc{0}\mybarc{3}\mybarc{0}\mybarc{0}}\\ 
    
    \multicolumn{1}{p{6.6cm}}{Use of security, quality and health metrics~\cite{8587375}} &
    \multicolumn{1}{c}{\cellcolor{NextBlue!60}{3.5}} &
    \multicolumn{1}{c}{\cellcolor{NextBlue!20}{4.0}} &
    \multicolumn{1}{c}{\mybarc{1}\mybarc{1}\mybarc{5}\mybarc{8}\mybarc{2}} &
    \multicolumn{1}{c}{\cellcolor{NextBlue!60}{2.6}} &
    \multicolumn{1}{c}{\cellcolor{NextBlue!60}{3.0}} &
    \multicolumn{1}{c}{\mybarc{2}\mybarc{6}\mybarc{7}\mybarc{1}\mybarc{1}} &
    \multicolumn{1}{c}{ 1.35}& \multicolumn{1}{c}{Y\mybarc{9}\mybarc{21}N} &
    \multicolumn{1}{c}{\cellcolor{NextBlue!60}{2.7}}
    &  \multicolumn{1}{c}{\cellcolor{NextBlue!60}{3.0}} &
    \multicolumn{1}{c}{\mybarc{6}\mybarc{6}\mybarc{12}\mybarc{4}\mybarc{2}}\\ 
    
    \multicolumn{1}{p{6.6cm}}{Typo guard/Typo
    detection~\cite{defendingAgainstPackageTyposquatting,duan2020towards}}
    &  \multicolumn{1}{c}{\cellcolor{NextBlue!20}{3.9}} &
    \multicolumn{1}{c}{\cellcolor{NextBlue!20}{4.0}} &
    \multicolumn{1}{c}{\mybarc{0}\mybarc{2}\mybarc{3}\mybarc{6}\mybarc{6}} &
    \multicolumn{1}{c}{\cellcolor{NextBlue!60}{2.9}} &
    \multicolumn{1}{c}{\cellcolor{NextBlue!60}{4.0}} &
    \multicolumn{1}{c}{\mybarc{1}\mybarc{4}\mybarc{7}\mybarc{5}\mybarc{0}} &
    \multicolumn{1}{c}{ 1.34} & \multicolumn{1}{c}{Y\mybarc{3}\mybarc{22}N} &
    \multicolumn{1}{c}{\cellcolor{NextBlue!60}{3.1}} &
    \multicolumn{1}{c}{\cellcolor{NextBlue!60}{3.0}} &
    \multicolumn{1}{c}{\mybarc{1}\mybarc{7}\mybarc{9}\mybarc{3}\mybarc{4}}\\ 
    
    \multicolumn{1}{p{6.6cm}}{Use minimal set of trusted build dependencies in the
    release job~\cite{10.1145/2491055.2491070}} &
    \multicolumn{1}{c}{\cellcolor{NextBlue!20}{4.1}}&
    \multicolumn{1}{c}{\cellcolor{NextBlue!20}{4.0}} &
    \multicolumn{1}{c}{\mybarc{0}\mybarc{1}\mybarc{3}\mybarc{6}\mybarc{7}} &
    \multicolumn{1}{c}{\cellcolor{NextBlue!60}{3.1}}&
    \multicolumn{1}{c}{\cellcolor{NextBlue!60}{3.0}} &
    \multicolumn{1}{c}{\mybarc{2}\mybarc{3}\mybarc{6}\mybarc{4}\mybarc{2}}&
    \multicolumn{1}{c}{ 1.32}& \multicolumn{1}{c}{Y\mybarc{2}\mybarc{2}N} &
    \multicolumn{1}{c}{\cellcolor{NextBlue!20}{3.8}} &
    \multicolumn{1}{c}{\cellcolor{NextBlue!20}{4.0}} &
    \multicolumn{1}{c}{\mybarc{0}\mybarc{3}\mybarc{5}\mybarc{3}\mybarc{6}}\\ 
    
    \multicolumn{1}{p{6.6cm}}{Integrity check of dependencies through
    cryptographic hashes~\cite{slsaframework,5402538,8807537,236322,203201,10.5555/2930611.2930648,alhamed2013security}} &
    \multicolumn{1}{c}{\cellcolor{NextBlue!60}{3.3}} &
    \multicolumn{1}{c}{\cellcolor{NextBlue!60}{3.0}} &
    \multicolumn{1}{c}{\mybarc{0}\mybarc{5}\mybarc{5}\mybarc{4}\mybarc{3}} &
    \multicolumn{1}{c}{\cellcolor{NextBlue}{2.5}}&
    \multicolumn{1}{c}{\cellcolor{NextBlue}{2.0}} &
    \multicolumn{1}{c}{\mybarc{3}\mybarc{6}\mybarc{6}\mybarc{0}\mybarc{2}}  &
    \multicolumn{1}{c}{1.32} & \multicolumn{1}{c}{Y\mybarc{12}\mybarc{10}N} &
    \multicolumn{1}{c}{\cellcolor{NextBlue}{2.3}}
    &  \multicolumn{1}{c}{\cellcolor{NextBlue}{2.0}} &
    \multicolumn{1}{c}{\mybarc{5}\mybarc{9}\mybarc{5}\mybarc{3}\mybarc{0}}\\ 
    
    \multicolumn{1}{p{6.6cm}}{Maintain detailed
    \ac{SBOM}~\cite{deliveruncompromised,herr2021breaking,forsgren20212020,Du2013TowardsAA,6107848} and perform \ac{SCA}~\cite{deliveruncompromised,10.1145/3459960.3459973,10.1145/3475716.3475769,zhang2015assessing,du2013towards,6107848,5929123,5428501}} &
    \multicolumn{1}{c}{\cellcolor{NextBlue!20}{4.2}} &
    \multicolumn{1}{c}{\cellcolor{NextBlue!20}{5.0}} &
    \multicolumn{1}{c}{\mybarc{0}\mybarc{2}\mybarc{4}\mybarc{4}\mybarc{9}}  &
    \multicolumn{1}{c}{\cellcolor{NextBlue!60}{3.4}} &
    \multicolumn{1}{c}{\cellcolor{NextBlue!20}{4.0}} &
    \multicolumn{1}{c}{\mybarc{1}\mybarc{3}\mybarc{4}\mybarc{7}\mybarc{2}} &
    \multicolumn{1}{c}{ 1.24 }& \multicolumn{1}{c}{Y\mybarc{15}\mybarc{16}N} &
    \multicolumn{1}{c}{\cellcolor{NextBlue!60}{2.9}}
    &  \multicolumn{1}{c}{\cellcolor{NextBlue!60}{3.0}} &
    \multicolumn{1}{c}{\mybarc{2}\mybarc{9}\mybarc{12}\mybarc{4}\mybarc{3}}\\ 
    
    \multicolumn{1}{p{6.6cm}}{Ephemeral build environment~\cite{10.1145/2491055.2491070,slsaframework}} &
    \multicolumn{1}{c}{\cellcolor{NextBlue!20}{3.6}}&
    \multicolumn{1}{c}{\cellcolor{NextBlue!60}{3.0}} &
    \multicolumn{1}{c}{\mybarc{0}\mybarc{3}\mybarc{6}\mybarc{3}\mybarc{5}} &
    \multicolumn{1}{c}{\cellcolor{NextBlue!60}{2.9}}&
    \multicolumn{1}{c}{\cellcolor{NextBlue!60}{3.0}} &
    \multicolumn{1}{c}{\mybarc{3}\mybarc{3}\mybarc{6}\mybarc{3}\mybarc{2}} &
    \multicolumn{1}{c}{ 1.24}& \multicolumn{1}{c}{Y\mybarc{3}\mybarc{1}N} &
    \multicolumn{1}{c}{\cellcolor{NextBlue!60}{2.8}} &
    \multicolumn{1}{c}{\cellcolor{NextBlue}{2.5}} &
    \multicolumn{1}{c}{\mybarc{1}\mybarc{1}\mybarc{1}\mybarc{0}\mybarc{1}}\\ 
    
    \multicolumn{1}{p{6.6cm}}{Prevent script execution} &
    \multicolumn{1}{c}{\cellcolor{NextBlue!20}{3.7}} &
    \multicolumn{1}{c}{\cellcolor{NextBlue!60}{3.0}} &
    \multicolumn{1}{c}{\mybarc{0}\mybarc{0}\mybarc{9}\mybarc{4}\mybarc{4}} &
    \multicolumn{1}{c}{\cellcolor{NextBlue!60}{3.0}} &
    \multicolumn{1}{c}{\cellcolor{NextBlue!60}{3.0}} &
    \multicolumn{1}{c}{\mybarc{2}\mybarc{3}\mybarc{7}\mybarc{3}\mybarc{2}} &
    \multicolumn{1}{c}{1.23}& \multicolumn{1}{c}{Y\mybarc{8}\mybarc{15}N} &
    \multicolumn{1}{c}{\cellcolor{NextBlue}{2.4}} &
    \multicolumn{1}{c}{\cellcolor{NextBlue}{2.0}} &
    \multicolumn{1}{c}{\mybarc{6}\mybarc{7}\mybarc{6}\mybarc{2}\mybarc{2}}\\ 
    
    \multicolumn{1}{p{6.6cm}}{Pull/Merge request review~\cite{goyal2018identifying}}  &
    \multicolumn{1}{c}{\cellcolor{NextBlue!20} 4.6} &
    \multicolumn{1}{c}{\cellcolor{NextBlue!20} 5.0}&
    \multicolumn{1}{c}{\mybarc{0}\mybarc{0}\mybarc{1}\mybarc{5}\mybarc{11}} &
    \multicolumn{1}{c}{\cellcolor{NextBlue!20}3.8}&
    \multicolumn{1}{c}{\cellcolor{NextBlue!20}4.0} &
    \multicolumn{1}{c}{\mybarc{0}\mybarc{1}\mybarc{6}\mybarc{6}\mybarc{4}} &
    \multicolumn{1}{c}{ 1.21}& \multicolumn{1}{c}{Y\mybarc{5}\mybarc{0}N} &
    \multicolumn{1}{c}{\cellcolor{NextBlue!20} 3.6} &
    \multicolumn{1}{c}{\cellcolor{NextBlue!20} 4.0} &
    \multicolumn{1}{c}{\mybarc{0}\mybarc{1}\mybarc{1}\mybarc{2}\mybarc{1}}\\ 
    
    \multicolumn{1}{p{6.6cm}}{Restrict access to system resources of code executed
    during each build steps~\cite{10.1145/2491055.2491070,vasilakis2018breakapp,10.1145/3144555.3144562}} &
    \multicolumn{1}{c}{\cellcolor{NextBlue!20} 4.0}&
    \multicolumn{1}{c}{\cellcolor{NextBlue!20} 4.0} &
    \multicolumn{1}{c}{\mybarc{0}\mybarc{1}\mybarc{3}\mybarc{8}\mybarc{5}} &
    \multicolumn{1}{c}{\cellcolor{NextBlue!60} 3.3}&
    \multicolumn{1}{c}{\cellcolor{NextBlue!60} 3.0} &
    \multicolumn{1}{c}{\mybarc{0}\mybarc{4}\mybarc{5}\mybarc{7}\mybarc{1}}&
    \multicolumn{1}{c}{ 1.21}& \multicolumn{1}{c}{Y\mybarc{0}\mybarc{4}N} &
    \multicolumn{1}{c}{\cellcolor{NextBlue!20} 3.8} &
    \multicolumn{1}{c}{\cellcolor{NextBlue!60} 3.5} &
    \multicolumn{1}{c}{\mybarc{0}\mybarc{0}\mybarc{2}\mybarc{1}\mybarc{1}}\\ 
    
    \multicolumn{1}{p{6.6cm}}{Code signing~\cite{236322,203201,10.5555/2930611.2930648,7180277,alhamed2013security,10.1145/1866307.1866315,10.1145/1137627.1137636}}&
    \multicolumn{1}{c}{\cellcolor{NextBlue!20} 3.7} &
    \multicolumn{1}{c}{\cellcolor{NextBlue!20} 4.0} &
    \multicolumn{1}{c}{\mybarc{0}\mybarc{1}\mybarc{7}\mybarc{5}\mybarc{4}} &
    \multicolumn{1}{c}{\cellcolor{NextBlue!60} 3.1} &
    \multicolumn{1}{c}{\cellcolor{NextBlue!60} 3.0} &
    \multicolumn{1}{c}{\mybarc{3}\mybarc{1}\mybarc{6}\mybarc{5}\mybarc{2}} &
    \multicolumn{1}{c}{ 1.19}& \multicolumn{1}{c}{Y\mybarc{11}\mybarc{19}N} &
    \multicolumn{1}{c}{\cellcolor{NextBlue!60} 3.1}
    &  \multicolumn{1}{c}{\cellcolor{NextBlue!60} 3.0} &
    \multicolumn{1}{c}{\mybarc{4}\mybarc{8}\mybarc{8}\mybarc{2}\mybarc{8}}\\ 
    
    \multicolumn{1}{p{6.6cm}}{Integrate Open-Source vulnerability scanner into
    CI/CD pipeline} &  \multicolumn{1}{c}{\cellcolor{NextBlue!20} 3.8} &
    \multicolumn{1}{c}{\cellcolor{NextBlue!20} 4.0} &
    \multicolumn{1}{c}{\mybarc{1}\mybarc{0}\mybarc{6}\mybarc{5}\mybarc{5}} &
    \multicolumn{1}{c}{\cellcolor{NextBlue!60} 3.3} &
    \multicolumn{1}{c}{\cellcolor{NextBlue!60} 3.0} &
    \multicolumn{1}{c}{\mybarc{1}\mybarc{1}\mybarc{8}\mybarc{6}\mybarc{1}} &
    \multicolumn{1}{c}{1.15}& \multicolumn{1}{c}{Y\mybarc{18}\mybarc{4}N} &
    \multicolumn{1}{c}{\cellcolor{NextBlue!60} 3.1}
    &  \multicolumn{1}{c}{\cellcolor{NextBlue!60} 3.0} &
    \multicolumn{1}{c}{\mybarc{1}\mybarc{6}\mybarc{7}\mybarc{5}\mybarc{3}}\\ 
    
    \multicolumn{1}{p{6.6cm}}{Use of dedicated build service~\cite{slsaframework}}
    &  \multicolumn{1}{c}{\cellcolor{NextBlue!20} 3.6}&
    \multicolumn{1}{c}{\cellcolor{NextBlue!20} 4.0} &
    \multicolumn{1}{c}{\mybarc{0}\mybarc{1}\mybarc{7}\mybarc{6}\mybarc{3}} &
    \multicolumn{1}{c}{\cellcolor{NextBlue!60} 3.3} &
    \multicolumn{1}{c}{\cellcolor{NextBlue!60} 3.0}&
    \multicolumn{1}{c}{\mybarc{1}\mybarc{2}\mybarc{7}\mybarc{4}\mybarc{3}}&
    \multicolumn{1}{c}{ 1.09}& \multicolumn{1}{c}{Y\mybarc{3}\mybarc{1}N} &
    \multicolumn{1}{c}{\cellcolor{NextBlue!60} 3.0} &
    \multicolumn{1}{c}{\cellcolor{NextBlue!60} 3.0} &
    \multicolumn{1}{c}{\mybarc{1}\mybarc{0}\mybarc{2}\mybarc{0}\mybarc{1}}\\ 
    
    \multicolumn{1}{p{6.6cm}}{Preventive squatting the released packages} &
    \multicolumn{1}{c}{\cellcolor{NextBlue!60} 3.1} &
    \multicolumn{1}{c}{\cellcolor{NextBlue!60} 3.0} &
    \multicolumn{1}{c}{\mybarc{2}\mybarc{2}\mybarc{8}\mybarc{2}\mybarc{3}} &
    \multicolumn{1}{c}{\cellcolor{NextBlue!60} 2.9}&
    \multicolumn{1}{c}{\cellcolor{NextBlue!60} 3.0} &
    \multicolumn{1}{c}{\mybarc{3}\mybarc{4}\mybarc{5}\mybarc{2}\mybarc{3}}  &
    \multicolumn{1}{c}{ 1.07}& \multicolumn{1}{c}{Y\mybarc{0}\mybarc{4}N} &
    \multicolumn{1}{c}{\cellcolor{NextBlue!20} 3.8} &
    \multicolumn{1}{c}{\cellcolor{NextBlue!60} 3.5} &
    \multicolumn{1}{c}{\mybarc{0}\mybarc{0}\mybarc{2}\mybarc{1}\mybarc{1}}\\  
    
    \multicolumn{1}{p{6.6cm}}{Audit, security assessment, vulnerability
    assessment, penetration testing} &
    \multicolumn{1}{c}{\cellcolor{NextBlue!20} 4.3} &
    \multicolumn{1}{c}{\cellcolor{NextBlue!20} 4.0} &
    \multicolumn{1}{c}{\mybarc{0}\mybarc{0}\mybarc{2}\mybarc{7}\mybarc{8}} &
    \multicolumn{1}{c}{\cellcolor{NextBlue!20} 4.1} &
    \multicolumn{1}{c}{\cellcolor{NextBlue!20} 4.0} &
    \multicolumn{1}{c}{\mybarc{1}\mybarc{1}\mybarc{1}\mybarc{6}\mybarc{8}} &
    \multicolumn{1}{c}{ 1.05} & \multicolumn{1}{c}{Y\mybarc{3}\mybarc{1}N} &
    \multicolumn{1}{c}{\cellcolor{NextBlue!20} 3.8} &
    \multicolumn{1}{c}{\cellcolor{NextBlue!60} 3.5} &
    \multicolumn{1}{c}{\mybarc{0}\mybarc{0}\mybarc{1}\mybarc{2}\mybarc{1}}\\  
    
    \multicolumn{1}{p{6.6cm}}{Reproducible builds~\cite{reprodbuilds,9240695,10.1145/2664243.2664288}}  &
    \multicolumn{1}{c}{\cellcolor{NextBlue!20} 4.2} &
    \multicolumn{1}{c}{\cellcolor{NextBlue!20} 5.0} &
    \multicolumn{1}{c}{\mybarc{0}\mybarc{1}\mybarc{5}\mybarc{1}\mybarc{10}} &
    \multicolumn{1}{c}{\cellcolor{NextBlue!20} 4.1} &
    \multicolumn{1}{c}{\cellcolor{NextBlue!20} 4.0} &
    \multicolumn{1}{c}{\mybarc{0}\mybarc{1}\mybarc{4}\mybarc{5}\mybarc{7}} &
    \multicolumn{1}{c}{ 1.02}& \multicolumn{1}{c}{Y\mybarc{8}\mybarc{22}N} &
    \multicolumn{1}{c}{\cellcolor{NextBlue!60} 3.5}
    &  \multicolumn{1}{c}{\cellcolor{NextBlue!20} 4.0} &
    \multicolumn{1}{c}{\mybarc{2}\mybarc{2}\mybarc{8}\mybarc{14}\mybarc{4}}\\ 
    
    \multicolumn{1}{p{6.6cm}}{Isolation of build
    steps~\cite{10.1145/2491055.2491070}} &
    \multicolumn{1}{c}{\cellcolor{NextBlue!60} 3.1} &
    \multicolumn{1}{c}{\cellcolor{NextBlue!60} 3.0} &
    \multicolumn{1}{c}{\mybarc{1}\mybarc{2}\mybarc{9}\mybarc{4}\mybarc{1}} &
    \multicolumn{1}{c}{\cellcolor{NextBlue!60} 3.1}&
    \multicolumn{1}{c}{\cellcolor{NextBlue!60} 3.0} &
    \multicolumn{1}{c}{\mybarc{1}\mybarc{3}\mybarc{7}\mybarc{5}\mybarc{1}} &
    \multicolumn{1}{c}{ 1.00} & \multicolumn{1}{c}{Y\mybarc{3}\mybarc{2}N} &
    \multicolumn{1}{c}{\cellcolor{NextBlue!60} 3.2} &
    \multicolumn{1}{c}{\cellcolor{NextBlue!60} 3.0} &
    \multicolumn{1}{c}{\mybarc{0}\mybarc{2}\mybarc{1}\mybarc{1}\mybarc{1}}\\ 
    
    \multicolumn{1}{p{6.6cm}}{Scoped
    packages~\cite{dependencyconfusionschibsted,microsoftwhitepaper}} &
    \multicolumn{1}{c}{\cellcolor{NextBlue!60} 2.9} &
    \multicolumn{1}{c}{\cellcolor{NextBlue!60} 3.0} &
    \multicolumn{1}{c}{\mybarc{1}\mybarc{6}\mybarc{5}\mybarc{3}\mybarc{2}} &
    \multicolumn{1}{c}{\cellcolor{NextBlue!60} 2.9}&
    \multicolumn{1}{c}{\cellcolor{NextBlue!60} 3.0} &
    \multicolumn{1}{c}{\mybarc{1}\mybarc{5}\mybarc{7}\mybarc{3}\mybarc{1}} &
    \multicolumn{1}{c}{ 1.00} & \multicolumn{1}{c}{Y\mybarc{3}\mybarc{1}N} &
    \multicolumn{1}{c}{\cellcolor{NextBlue!60} 2.8} &
    \multicolumn{1}{c}{\cellcolor{NextBlue}{2.0}} &
    \multicolumn{1}{c}{\mybarc{0}\mybarc{3}\mybarc{0}\mybarc{0}\mybarc{1}}\\ 
    
    \multicolumn{1}{p{6.6cm}}{Establish internal repository mirrors and reference
    one private feed, not multiple~\cite{microsoftwhitepaper}} &
    \multicolumn{1}{c}{\cellcolor{NextBlue!20} 3.6} &
    \multicolumn{1}{c}{\cellcolor{NextBlue!60} 3.0} &
    \multicolumn{1}{c}{\mybarc{0}\mybarc{2}\mybarc{8}\mybarc{2}\mybarc{5}} &
    \multicolumn{1}{c}{\cellcolor{NextBlue!20} 3.7} &
    \multicolumn{1}{c}{\cellcolor{NextBlue!20} 4.0} &
    \multicolumn{1}{c}{\mybarc{0}\mybarc{2}\mybarc{5}\mybarc{6}\mybarc{4}}&
    \multicolumn{1}{c}{0.97} & \multicolumn{1}{c}{Y\mybarc{14}\mybarc{8}N} &
    \multicolumn{1}{c}{\cellcolor{NextBlue!60} 2.7}
    &  \multicolumn{1}{c}{\cellcolor{NextBlue!60} 3.0} &
    \multicolumn{1}{c}{\mybarc{3}\mybarc{6}\mybarc{9}\mybarc{3}\mybarc{1}}\\ 
    
    \multicolumn{1}{p{6.6cm}}{\acl{AST}~\cite{9402087,kaplan2021survey,ohmbuildwatch,10.1145/3372297.3420015,ohm2020supporting,Garrett2019DetectingSP,8530068,10.1145/3239235.3268920,10.1145/3098954.3120928,5929123,5428501,sejfia2022practical}} &
    \multicolumn{1}{c}{\cellcolor{NextBlue!20} 4.1} &
    \multicolumn{1}{c}{\cellcolor{NextBlue!20} 4.0} &
    \multicolumn{1}{c}{\mybarc{0}\mybarc{0}\mybarc{6}\mybarc{4}\mybarc{7}} &
    \multicolumn{1}{c}{\cellcolor{NextBlue!20} 4.3} &
    \multicolumn{1}{c}{\cellcolor{NextBlue!20} 5.0} &
    \multicolumn{1}{c}{\mybarc{0}\mybarc{1}\mybarc{2}\mybarc{4}\mybarc{10}}&
    \multicolumn{1}{c}{ 0.95} & \multicolumn{1}{c}{Y\mybarc{10}\mybarc{15}N} &
    \multicolumn{1}{c}{\cellcolor{NextBlue!20} 3.7} &
    \multicolumn{1}{c}{\cellcolor{NextBlue!60} 3.0} &
    \multicolumn{1}{c}{\mybarc{0}\mybarc{1}\mybarc{12}\mybarc{4}\mybarc{7}}\\ 
    
    \multicolumn{1}{p{6.6cm}}{Establish vetting process for Open-Source components
    hosted in internal/public repositories~\cite{duan2020towards,10.1145/3460120.3484736,236368,Garrett2019DetectingSP,10.1145/3196398.3196401}} &
    \multicolumn{1}{c}{\cellcolor{NextBlue!20} 4.1} &
    \multicolumn{1}{c}{\cellcolor{NextBlue!20} 4.0} &
    \multicolumn{1}{c}{\mybarc{0}\mybarc{0}\mybarc{4}\mybarc{7}\mybarc{6}} &
    \multicolumn{1}{c}{\cellcolor{NextBlue!20} 4.3} &
    \multicolumn{1}{c}{\cellcolor{NextBlue!20} 5.0} &
    \multicolumn{1}{c}{\mybarc{0}\mybarc{1}\mybarc{1}\mybarc{6}\mybarc{9}} &
    \multicolumn{1}{c}{ 0.95} & \multicolumn{1}{c}{Y\mybarc{7}\mybarc{18}N} &
    \multicolumn{1}{c}{\cellcolor{NextBlue!20} 3.8} &
    \multicolumn{1}{c}{\cellcolor{NextBlue!60} 3.5} &
    \multicolumn{1}{c}{\mybarc{0}\mybarc{0}\mybarc{12}\mybarc{6}\mybarc{6}}\\ 
    
    \multicolumn{1}{p{6.6cm}}{Code isolation and sandboxing~\cite{9402108,vasilakis2018breakapp,10.1145/3144555.3144562}} &
    \multicolumn{1}{c}{\cellcolor{NextBlue!20} 3.9} &
    \multicolumn{1}{c}{\cellcolor{NextBlue!20} 4.0} &
    \multicolumn{1}{c}{\mybarc{0}\mybarc{1}\mybarc{5}\mybarc{6}\mybarc{5}} &
    \multicolumn{1}{c}{\cellcolor{NextBlue!20} 4.2} &
    \multicolumn{1}{c}{\cellcolor{NextBlue!20} 4.0} &
    \multicolumn{1}{c}{\mybarc{0}\mybarc{1}\mybarc{2}\mybarc{6}\mybarc{8}} &
    \multicolumn{1}{c}{0.93} & \multicolumn{1}{c}{Y\mybarc{4}\mybarc{18}N} &
    \multicolumn{1}{c}{\cellcolor{NextBlue!60} 3.2}
    &  \multicolumn{1}{c}{\cellcolor{NextBlue!60} 3.0} &
    \multicolumn{1}{c}{\mybarc{2}\mybarc{2}\mybarc{11}\mybarc{4}\mybarc{3}}\\ 
    
    \multicolumn{1}{p{6.6cm}}{\acl{RASP}} &
    \multicolumn{1}{c}{\cellcolor{NextBlue!20} 3.7} &
    \multicolumn{1}{c}{\cellcolor{NextBlue!20} 4.0} &
    \multicolumn{1}{c}{\mybarc{0}\mybarc{3}\mybarc{5}\mybarc{3}\mybarc{6}} &
    \multicolumn{1}{c}{\cellcolor{NextBlue!20} 4.2} &
    \multicolumn{1}{c}{\cellcolor{NextBlue!20} 4.0} &
    \multicolumn{1}{c}{\mybarc{0}\mybarc{2}\mybarc{0}\mybarc{7}\mybarc{8}}&
    \multicolumn{1}{c}{0.88} & \multicolumn{1}{c}{Y\mybarc{5}\mybarc{18}N} &
    \multicolumn{1}{c}{\cellcolor{NextBlue!20} 3.8}
    &  \multicolumn{1}{c}{\cellcolor{NextBlue!20} 4.0} &
    \multicolumn{1}{c}{\mybarc{0}\mybarc{1}\mybarc{6}\mybarc{11}\mybarc{4}}\\ 
    
    \multicolumn{1}{p{6.6cm}}{Manual source code review~\cite{kaplan2021survey}} &
    \multicolumn{1}{c}{\cellcolor{NextBlue!20} 4.1} &
    \multicolumn{1}{c}{\cellcolor{NextBlue!20} 4.0} &
    \multicolumn{1}{c}{\mybarc{0}\mybarc{2}\mybarc{1}\mybarc{7}\mybarc{7}} &
    \multicolumn{1}{c}{\cellcolor{NextBlue!20} 4.8} &
    \multicolumn{1}{c}{\cellcolor{NextBlue!20} 5.0} &
    \multicolumn{1}{c}{\mybarc{0}\mybarc{0}\mybarc{1}\mybarc{2}\mybarc{14}}&
    \multicolumn{1}{c}{ 0.85}& \multicolumn{1}{c}{Y\mybarc{5}\mybarc{20}N} &
    \multicolumn{1}{c}{\cellcolor{NextBlue!20} 4.4} &
    \multicolumn{1}{c}{\cellcolor{NextBlue!20} 5.0} &
    \multicolumn{1}{c}{\mybarc{1}\mybarc{0}\mybarc{3}\mybarc{5}\mybarc{15}}\\ 
    
    \multicolumn{1}{p{6.6cm}}{Build dependencies from sources} &
    \multicolumn{1}{c}{\cellcolor{NextBlue!60} 3.0} &
    \multicolumn{1}{c}{\cellcolor{NextBlue!60} 3.0} &
    \multicolumn{1}{c}{\mybarc{1}\mybarc{3}\mybarc{9}\mybarc{3}\mybarc{1}} &
    \multicolumn{1}{c}{\cellcolor{NextBlue!20} 4.1}&
    \multicolumn{1}{c}{\cellcolor{NextBlue!20} 4.0} &
    \multicolumn{1}{c}{\mybarc{1}\mybarc{0}\mybarc{4}\mybarc{4}\mybarc{8}} &
    \multicolumn{1}{c}{ 0.73} & \multicolumn{1}{c}{Y\mybarc{7}\mybarc{18}N} &
    \multicolumn{1}{c}{\cellcolor{NextBlue!20} 3.8} &
    \multicolumn{1}{c}{\cellcolor{NextBlue!20} 4.0} &
    \multicolumn{1}{c}{\mybarc{1}\mybarc{1}\mybarc{5}\mybarc{13}\mybarc{4}}\\ 
    
    \bottomrule

    \end{tabular}
    } \caption[]{Assessment of safeguards by 17 domain experts (left) and 134
    developers (right). Utility and cost assessments were given on a Likert
    scale, the numbers are shown with bar plots, from 1 (low) to 5 (high). The
    background of mean and median values are determined by the intervals
    \colorbox{NextBlue}{{$[1,2.5]$}}, \colorbox{NextBlue!60}{$(2.5,3.5]$} and
    \colorbox{NextBlue!20}{$(3.5,5.0]$}. Safeguards are
    shown in the order of the mean of their \textbf{Utility-to-Cost
    Ratio (U/C)} (descending). Developer feedback on safeguard use was
    collected with yes/no questions, the number of respective answers are shown
    using a bar plot.}
    \label{tab:safeguards2}
    \end{table*}

\section{Discussion}\label{section:discussion} While the taxonomy presented in
Section~\ref{section:attacktaxonomy} is largely agnostic to ecosystems, this
section discusses differences between ecosystems and highlights possible future 
research on the basis of our work.

\subsubsection{Differences between Ecosystems}

As mentioned in Section~\ref{section:attackermodel}, the attacker's high-level
goal is to \textbf{inject malicious code} into open-source artifacts such that it is
executed downstream. Several techniques to this end
are indeed independent of specific ecosystems/languages, e.g.,
\textit{Take-over Legitimate Account} or \textit{Become Maintainer}.

Other attack vectors, however, are specific: \textit{Abuse Dependency Resolution
Mechanism} attacks depend on the approach and strategy used by the respective
package manager to resolve and download declared dependencies from internal and
external repositories. For instance, Maven, npm, pip, NuGet or Composer were
affected by the dependency confusion attack, while Go and Cargo were
not~\cite{dependencyconfusionschibsted}. Several attacks below \textit{Exploit
Rendering Weakness} depend on the interpretation and visualization of (Unicode)
characters by user interfaces and
compiler/interpreters~\cite{boucher2021trojan}. Also name confusion attacks
need to consider ecosystem specificities, esp. \textit{Built-In Packages}.


More differences exist when it comes to the \textbf{execution or trigger of malicious
code}, which is beyond the taxonomy's primary focus on code injection. For Python
and Node.js, this is commonly achieved through installation hooks, which trigger
the execution of code provided in the downloaded package (e.g., in
\texttt{setup.py} for Python or \texttt{package.json} for JavaScript). A
comparable feature is not present in most compiled languages, like Java, C/C++
or Ruby. In such cases, execution is achieved either at runtime, e.g., by
embedding the payload in a specific function or initializer, or by poisoning
test routines~\cite{ohm2020backstabbers}.






Differences also exist in regards to \textbf{code obfuscation and malware detection}. In
case of interpreted languages, downloaded packages contain the malware's source
code, which makes it more accessible to analysts compared to compiled languages.
The presence of encoded or encrypted code in such packages proofed being a good
indicator of compromise~\cite{sejfia2022practical}, as there are few legimitate
use-cases for open-source packages. Minification is one of them, however,
matters primarily for frontend JavaScript libraries. Indeed, many existing
attacks did not employ obfuscation or encryption~\cite{ohm2020backstabbers}
techniques. Still, the quantity of open-source packages and versions makes
manual inspection very difficult, even if source code is accessible.

When it comes to compiled code, well-known techniques like packing, dead-code
insertion or subroutine reordering~\cite{5633410} make reverse engineering and
analysis more complex. It is also noteworthy that ecosystems for interpreted
languages ship compiled code. For instance, many Python libraries for ML/AI
use-cases include and wrap platform-specific C/C++ binaries.


For what concerns \textbf{safeguards}, several of them are specific to selected
package managers, namely \textit{Scoped packages} (Node.js) and \textit{Prevent
script execution} (Python and Node.js). All others are relevant no matter the
ecosystem, however, control implementations and technology choices differ,
e.g., in case of \textit{Application Security Testing}.
Duan et al.~\cite{duan2020towards} present a comparative framework for security
features of package repositories (exemplified with PyPI, npm and RubyGems).

\subsubsection{Benefits of the Taxonomy for Future Research}
Our work systematizes knowledge about \ac{OSS} supply chain security by
abstracting, contextualizing and classifying existing works. The proposed
taxonomy can benefit future research by offering a central point of reference
and a common terminology.

The comprehensive list of attack vectors and safeguards can support assessing
the security level of open-source projects, e.g., to conduct comparative empiric
studies across projects and ecosystems and over time.


The availability of source code in ecosystems for interpreted languages suggests
that malware analysis is more straight-forward. Still, recent publications focus
on those ecosystems, esp. JavaScript and
Python~\cite{duan2020towards,DBLP:journals/corr/abs-2102-06301,236368,sejfia2022practical,9402108}),
partly due to their popularity, but also because existing malware analysis
techniques cannot be easily applied.

When attacks involve the intentional insertion of vulnerabilities, for example,
it requires the analysis of the modification's context to distinguish it from a
vulnerability that has been accidentaly introduced~\cite{9402087}.

Also code generation and the difficulty to identify the VCS' commit
corresponding to a given pre-built package make malware analyis of source code
difficult. Works like~\cite{vu95py2src,10.1145/3468264.3468592} describe the
challenges to identify the discrepancy between source code in VCS and in
pre-built components.

The safeguard \textit{Reproducible builds} \cite{reprodbuilds,9240695,
10.1145/2664243.2664288} addresses this problem, however, it is not commonly
applied, considered costly (cf. Table \ref{tab:safeguards2}) and more complex
projects require significant implementation efforts.

\section{User Survey Demographics}\label{section:usersurveydemographics}

This section provides demographic information about the respondents of the two online surveys.
In summary, the respondents to the expert survey meet the requirement of being experts in the domain and
participate actively in \ac{OSS} projects. The respondents to the developer
survey regularly consume \ac{OSS} and have little knowledge of supply chain
security. 

\subsubsection*{Domain Experts}



\textbf{17 respondents} participated in the online survey designed for experts
in the domain of \emph{software supply chain security}. According to the
self-assessment of their skills, 12 respondents consider themselves
\textbf{knowledgeable in the domain of supply chain security}, but also in
software security (14) and development (12).
Considering their acquaintance with 11 popular languages~\cite{githuboctoverse2021}, 
the respondents cover 9 out of them, whereby Python, Java and JavaScript are covered 
best, while nobody had a background in .NET and Objective-C.
14 out of the 17 respondents are active participants in \ac{OSS} projects, and
were asked about their respective role (multiple choice): all 14 are
contributors, 7 are project maintainers, and 3 exercise other roles.
9 experts work in the private sector, compared to 5 working in the public sector
(e.g., government, academia) and 3 in the not-for-profit sector. They cover the
industry sectors information industry (8), computer industry (2),
telecommunications (2), entertainment industry (1), mass media (1), defense (1)
and others (2).

\subsubsection*{Developers}

\textbf{134 respondents} participated in the online survey designed for software
developers, who were assumed to exercise the role of downstream consumers in
\ac{OSS} supply chains. 
This assumption was confirmed given that 121 (90\%) responded 
to using open-source components in their daily job. Moreover, 37 (28\%) actively participate in
\ac{OSS} projects: 31 as contributors and 22 as maintainers (multiple choice).
74 are also maintainers of code repositories, and 21 administer package
repositories.
The self-assessment of their skills shows that they are \textbf{knowledgeable in
software development} (113), and less so in supply chain security (22) and
software security (44). 
They cover all of the 11 programming languages (multiple
choice), whereby Java,
JavaScript
and Python
are the most popular ones. 
The majority of the respondents (120) work in the private sector. In terms of
industry sectors, computer and information industry (55 and 54) outweigh
other sectors.

\section{Related Works}
\label{section:relatedworks}

In the following, we distinguish existing works related to specific aspects of
\ac{OSS} supply chains, e.g., technologies, systems, or stakeholder interactions,
from more general ones covering the entire supply chain. Both contributed to the
initial set of attack vectors and safeguards.


\textbf{Specific Works.} Giovanini et al.~\cite{giovanini2021leveraging}
leverage patterns in team dynamics to predict the susceptibility of \ac{OSS}
development teams to social engineering attacks.
Gonzalez et al.~\cite{9402087} describe attacks aiming to inject malicious code
in \ac{VCS}s via commits. They propose a rule-based anomaly detector that uses
commit logs and repository metadata to detect potentially malicious commits. In
the same direction, Goyal et al.~\cite{goyalUnusual} analyze collaborative
\ac{OSS} development, and highlight the problem of overwhelming information that
potentially results in maintainers accepting malicious merge requests.
Wheeler~\cite{1565233} describes the problem of code injection into software by
subverted compilers, and proposes \ac{DDC} to detect such attack. Within this
context, Lamb et al.~\cite{reprodbuilds} propose an approach to determine the
correspondence between binaries and the related source code through bit-for-bit
checks of build processes, while Ly et al.~\cite{10.1145/3468264.3468592}
analyzed the discrepancies between Python code in a projects' \ac{VCS} and
its distributed artifacts.
Gruhn et al. \cite{10.1145/2491055.2491070} analyze the security of \ac{CI}
systems, and identify web \ac{UI}s and build processes as the main sources of
malicious data. They propose a secure build server architecture, based on the
isolation of build processes through virtualization. 
Multiple works address common threats to package managers of different
ecosystems. Cappos et al.
\cite{10.1145/1455770.1455841,packagemanagementsecurity} identify possible
attack vectors related to a lack of proper signature management at the level of
packages and their metadata, some of which we considered below \emph{Distribute
malicious version of legitimate package}.
Zimmerman et al. \cite{236368} analyze security threats and associated risks in
the npm ecosystem, and define several metrics describing the downstream reach of
packages and maintainers, which allows identifying critical elements. Inversely,
they also measure the number of implicitly trusted upstream packages and
maintainers. Bagmar et al. \cite{DBLP:journals/corr/abs-2102-06301} performed
similar work for the PyPI ecosystem, and several of their vectors are subsumed
below \textit{Create name confusion with legit. package}.
Duan et al. \cite{duan2020towards} propose a framework to qualitatively assess
functional and security aspects of package managers for interpreted languages
(i.e., Python, JavaScript, and Ruby). They provide an overview of stakeholders
(and their relationships) in those package manager ecosystems, but do not
specifically cover \ac{VCS} and build systems.
Also Kaplan et al. \cite{kaplan2021survey} present the state of the art of
threats in package repositories and describe \textendash~also experimental
\textendash~countermeasures from the scientific literature.

\textbf{General works.}
Ohm et al. \cite{ohm2020backstabbers} manually inspect malicious npm, PyPi, and
Ruby packages. They propose an attack tree \textendash~based on a graph of
Pfretzschner et al. \cite{10.1145/3098954.3120928} \textendash~describing how to
inject malicious code into dependency trees. The attack tree proposed in
Section~\ref{section:attacktaxonomy} follows a more rigorous structure (degrees
of interference with existing packages, supply chain stages, stakeholders and
systems involved) and our \ac{SLR} resulted in the addition of 89
attack vectors. Our results have been validated through two user surveys.
Du et al.~\cite{Du2013TowardsAA} describe a wide range of high-level software
supply chain risks, both external (e.g., natural disaster, political factor) and
internal ones (e.g., participants, software components).
ENISA~\cite{ENISAThreatSupplyChain} proposes a taxonomy of supply chain attacks
describing the techniques used by attackers and the targeted assets, both from the 
supplier and customer perspective. However, they only mention few
high-level techniques.
Torres-Arias et al. \cite{236322} propose in-toto, a framework based on the
concepts of delegations and roles to cryptographically ensure the integrity of
software supply chains through an end-to-end verification of each step and
actors involved. Samuel et al. \cite{10.1145/1866307.1866315} propose \ac{TUF}
to overcome in-toto's main limitations regarding secure distribution,
revocation and replacement of keys.  

\section{Threats to Validity}
\label{section:threatovalidity}

The taxonomy was modeled using the semantics of attack trees, and several of its
nodes reflect the characterizing stages of \ac{OSS} supply chains, with code
from project contributors and maintainers flowing to downstream consumers.
Though its comprehensiveness, comprehensibility, and usefulness have been
positively assessed by the survey participants, the taxonomy reflects the current
state of the art. As the supply chain technologies evolve, it is expected that 
the proposed attack tree will evolve too.

We systematically reviewed the literature and continuously monitor aggregators
of security news to create a comprehensive list of attack vectors, and
collected feedback from domain experts to assess its completeness. Still, the
complexity of \ac{OSS} supply chains makes it very likely that new attack
vectors and techniques will be discovered. The quality of the taxonomy will
correspond to the degree of changes required to reflect such new attacks.

\hp{Maybe we say this somewhere in the methodology already and remove it here?}

The feedback collected from survey participants could have been biased if we
only considered experts that we directly know. Instead, thanks to the snowball
sampling we have reached also people outside of our network. Considering
authors of relevant scientific works, experts from academia and industry, all
working in the specific area of software supply-chain security, allowed us to
reach the intended audience (cf. Section~\ref{section:usersurveydemographics}):
The 17 respondents of the expert survey were knowledgeable in supply chain
security and actively participate in \ac{OSS} projects, the 134 participants of
the developer survey have knowledge in software development and use \ac{OSS}
regularly, and both groups cover a diverse range of prg. languages, incl. those
subject to frequent attacks.



\section{Conclusion}
\label{section:conclusion}
As validated by domain experts, the proposed taxonomy of attacks on \ac{OSS}
supply chains is comprehensive, comprehensible, and serves different use-cases.
It can benefit future research serving as a central reference point and setting a
common terminology. 

The listing of safeguards and their mapping to attack tree nodes helps to
determine the exposure of given stakeholders to supply chain attacks. Their
assessment in terms of utility and costs can serve to optimize the spending of
limited security budgets. Future empiric studies may investigate the prevalence
of the identified countermeasures, e.g. their use by given open-source projects.
On our side, we aim at developing techniques for the detection of malicious code
in compiled Java open-source components.
Going forward, to raise awareness for threats to \ac{OSS} supply chains, we will
publish the interactive visualization of the taxonomy online. References to
literature and real-world incidents will be kept up-to-date by using the
open-source approach to help the taxonomy itself stay relevant.
Finally, we would also like to put the taxonomy into practice for other
use-cases, esp. risk assessment.

\small\noindent\textbf{Acknowledgements.}
We would like to thank all survey participants for their time and the
constructive and insightful feedback.
This work is partly funded by EU grant No.
830892 (SPARTA).
\normalsize

\bibliography{bibliography}

\begin{thebibliography}{100}
\bibitem{sharma}
A.~Sharma, ``Inside the "fallguys" malware that steals your browsing data and
  gaming ims; continued attack on open source software.''
\end{thebibliography}
\bibliographystyle{ieeetr}

\begin{appendices}

 \section{Safeguards against \ac{OSS} Supply Chain
 Attacks}\label{appendix:listofsafeguards} Table~\ref{tab:safeguards} shows the
 identified safeguards mitigating attacks on \ac{OSS} Supply Chain. 

 \begin{table*}[!hbtp]

    \centering
    \begin{tabular}{r|lllllllll|}
    \toprule
    
    \multicolumn{1}{l}{} & \multicolumn{4}{c}{\textbf{Control Type}} & \multicolumn{3}{c}{\textbf{Stakeholders Involved}} & \multicolumn{1}{l}{}\\ \cmidrule(lr){2-5} \cmidrule(lr){6-8}
    
    \multicolumn{1}{c}{\multirow{-2}{*}{\textbf{Safeguard}}} & \multicolumn{1}{l}{\rotatebox[origin=c]{70}{\textbf{Directive}}} & \multicolumn{1}{l}{\rotatebox[origin=c]{70}{\textbf{Preventive}}} & \multicolumn{1}{l}{\rotatebox[origin=c]{70}{\textbf{Detective}}} & \multicolumn{1}{l}{\rotatebox[origin=c]{70}{\textbf{Corrective}}} & \multicolumn{1}{c}{\rotatebox[origin=c]{70}{\textbf{\ac{OSS} Maintainer}}} & \multicolumn{1}{c}{\rotatebox[origin=c]{70}{\textbf{3P Service Prov.}}} & \multicolumn{1}{c}{\rotatebox[origin=c]{70}{\textbf{\ac{OSS} Consumer}}}& \multicolumn{1}{c}{\multirow{-2}{*}{\textbf{Attack-Vector Addressed}}} \\ 
    
    \midrule

    \multicolumn{1}{p{5.4cm}}{Maintain detailed \ac{SBOM} and perform \ac{SCA}} & \multicolumn{1}{l}{} & \multicolumn{1}{c}{\checkmark} & \multicolumn{1}{c}{\checkmark} & \multicolumn{1}{l}{} & \multicolumn{1}{c}{$\bullet$}& \multicolumn{1}{c}{$\bullet$}& \multicolumn{1}{c}{$\bullet$} &  \multicolumn{1}{c}{AV-000}\\ 
    \multicolumn{1}{p{5.4cm}}{Code signing} & \multicolumn{1}{c}{} & \multicolumn{1}{c}{} & \multicolumn{1}{c}{\checkmark} & \multicolumn{1}{l}{}& \multicolumn{1}{c}{$\bullet$}& \multicolumn{1}{c}{$\bullet$}& \multicolumn{1}{c}{$\bullet$} &  \multicolumn{1}{c}{AV-200, AV-500} \\ 
    \multicolumn{1}{p{5.4cm}}{Use of security, quality and health metrics} & \multicolumn{1}{l}{\checkmark} & \multicolumn{1}{l}{\checkmark} & \multicolumn{1}{l}{} & \multicolumn{1}{l}{}& \multicolumn{1}{c}{$\bullet$}& \multicolumn{1}{c}{$\bullet$}& \multicolumn{1}{c}{$\bullet$} &  \multicolumn{1}{c}{AV-000}\\ 
    \multicolumn{1}{p{5.4cm}}{Reproducible builds} & \multicolumn{1}{l}{} & \multicolumn{1}{l}{} & \multicolumn{1}{c}{\checkmark} & \multicolumn{1}{l}{}& \multicolumn{1}{c}{$\bullet$}& \multicolumn{1}{c}{$\bullet$}& \multicolumn{1}{c}{$\bullet$} & \multicolumn{1}{c}{AV-400, AV-500}\\ 

    \multicolumn{1}{p{5.4cm}}{Secure authentication (e.g., \ac{MFA}, password recycle, session timeout, token protection)} & \multicolumn{1}{c}{} & \multicolumn{1}{c}{\checkmark} & \multicolumn{1}{c}{} & \multicolumn{1}{c}{}& \multicolumn{1}{c}{$\bullet$}& \multicolumn{1}{c}{$\bullet$}& \multicolumn{1}{c}{} &  \multicolumn{1}{c}{AV-*00 $\rightarrow$ AV-602}\\ 
    \multicolumn{1}{p{5.4cm}}{User account management} & \multicolumn{1}{c}{} & \multicolumn{1}{c}{\checkmark} & \multicolumn{1}{c}{} & \multicolumn{1}{c}{\checkmark}& \multicolumn{1}{c}{$\bullet$}& \multicolumn{1}{c}{$\bullet$}& \multicolumn{1}{c}{}  &  \multicolumn{1}{c}{AV-302,AV-402,AV-504,AV-600}\\ 
    \multicolumn{1}{p{5.4cm}}{Audit} & \multicolumn{1}{c}{\checkmark} & \multicolumn{1}{c}{} & \multicolumn{1}{c}{\checkmark} & \multicolumn{1}{c}{}& \multicolumn{1}{c}{$\bullet$}& \multicolumn{1}{c}{$\bullet$}& \multicolumn{1}{c}{}  &  \multicolumn{1}{c}{AV-000}\\ 
    \multicolumn{1}{p{5.4cm}}{Security assessment} & \multicolumn{1}{c}{} & \multicolumn{1}{c}{} & \multicolumn{1}{c}{\checkmark} & \multicolumn{1}{c}{}& \multicolumn{1}{c}{$\bullet$}& \multicolumn{1}{c}{$\bullet$}& \multicolumn{1}{c}{}  &  \multicolumn{1}{c}{AV-000}\\ 
    \multicolumn{1}{p{5.4cm}}{Vulnerability assessment} & \multicolumn{1}{c}{} & \multicolumn{1}{c}{} & \multicolumn{1}{c}{\checkmark} & \multicolumn{1}{c}{} & \multicolumn{1}{c}{$\bullet$}& \multicolumn{1}{c}{$\bullet$}& \multicolumn{1}{c}{} &  \multicolumn{1}{c}{AV-000}\\ 
    \multicolumn{1}{p{5.4cm}}{Penetration testing} & \multicolumn{1}{c}{} & \multicolumn{1}{c}{} & \multicolumn{1}{c}{\checkmark} & \multicolumn{1}{c}{}& \multicolumn{1}{c}{$\bullet$}& \multicolumn{1}{c}{$\bullet$}& \multicolumn{1}{c}{}  &  \multicolumn{1}{c}{AV-000}\\ 
    \multicolumn{1}{p{5.4cm}}{Scoped packages} & \multicolumn{1}{c}{} & \multicolumn{1}{c}{\checkmark} & \multicolumn{1}{c}{} & \multicolumn{1}{c}{}& \multicolumn{1}{c}{$\bullet$}& \multicolumn{1}{c}{$\bullet$}& \multicolumn{1}{c}{}  &  \multicolumn{1}{c}{AV-509}\\ 
    \multicolumn{1}{p{5.4cm}}{Preventive squatting the released packages} & \multicolumn{1}{c}{} & \multicolumn{1}{c}{\checkmark} & \multicolumn{1}{c}{} & \multicolumn{1}{c}{}& \multicolumn{1}{c}{$\bullet$}& \multicolumn{1}{c}{$\bullet$}& \multicolumn{1}{c}{}  &  \multicolumn{1}{c}{AV-200}\\ 
    \multicolumn{1}{p{5.4cm}}{Pull/Merge request review} & \multicolumn{1}{c}{} & \multicolumn{1}{c}{\checkmark} & \multicolumn{1}{c}{} & \multicolumn{1}{c}{} & \multicolumn{1}{c}{$\bullet$}& \multicolumn{1}{c}{}& \multicolumn{1}{c}{}&  \multicolumn{1}{c}{AV-301, AV-302} \\ 
    \multicolumn{1}{p{5.4cm}}{Protect production branch} & \multicolumn{1}{c}{} & \multicolumn{1}{c}{\checkmark} & \multicolumn{1}{c}{\checkmark} & \multicolumn{1}{c}{}& \multicolumn{1}{c}{$\bullet$}& \multicolumn{1}{c}{}& \multicolumn{1}{c}{} &  \multicolumn{1}{c}{AV-301, AV-302} \\ 
    \multicolumn{1}{p{5.4cm}}{Isolation of build steps} & \multicolumn{1}{c}{} & \multicolumn{1}{c}{\checkmark} & \multicolumn{1}{c}{} & \multicolumn{1}{c}{}& \multicolumn{1}{c}{$\bullet$}& \multicolumn{1}{c}{}& \multicolumn{1}{c}{} &  \multicolumn{1}{c}{AV-400} \\ 
    \multicolumn{1}{p{5.4cm}}{Ephemeral build environment} & \multicolumn{1}{c}{} & \multicolumn{1}{c}{\checkmark} & \multicolumn{1}{c}{} & \multicolumn{1}{c}{}& \multicolumn{1}{c}{$\bullet$}& \multicolumn{1}{c}{}& \multicolumn{1}{c}{} &  \multicolumn{1}{c}{AV-400} \\ 
    \multicolumn{1}{p{5.4cm}}{Use minimal set of trusted build dependencies in the release job} & \multicolumn{1}{c}{} & \multicolumn{1}{c}{\checkmark} & \multicolumn{1}{c}{} & \multicolumn{1}{c}{}& \multicolumn{1}{c}{$\bullet$}& \multicolumn{1}{c}{}& \multicolumn{1}{c}{} &  \multicolumn{1}{c}{AV-400} \\ 
    \multicolumn{1}{p{5.4cm}}{Restrict access to system resources of code executed during each build steps} & \multicolumn{1}{c}{} & \multicolumn{1}{c}{\checkmark} & \multicolumn{1}{c}{} & \multicolumn{1}{c}{}& \multicolumn{1}{c}{$\bullet$}& \multicolumn{1}{c}{}& \multicolumn{1}{c}{} &  \multicolumn{1}{c}{AV-400} \\ 
    \multicolumn{1}{p{5.4cm}}{Use of dedicated build service} & \multicolumn{1}{c}{} & \multicolumn{1}{c}{\checkmark} & \multicolumn{1}{c}{} & \multicolumn{1}{c}{}& \multicolumn{1}{c}{$\bullet$}& \multicolumn{1}{c}{}& \multicolumn{1}{c}{} &  \multicolumn{1}{c}{AV-400 $\rightarrow$ AV-700 } \\ 


    \multicolumn{1}{p{5.4cm}}{Manual source code review} & \multicolumn{1}{c}{} & \multicolumn{1}{c}{} & \multicolumn{1}{c}{\checkmark} & \multicolumn{1}{c}{}& \multicolumn{1}{c}{}& \multicolumn{1}{c}{$\bullet$}& \multicolumn{1}{c}{$\bullet$}  &  \multicolumn{1}{c}{AV-300}\\ 
    \multicolumn{1}{p{5.4cm}}{Application Security Testing } & \multicolumn{1}{c}{} & \multicolumn{1}{c}{} & \multicolumn{1}{c}{\checkmark} & \multicolumn{1}{c}{}& \multicolumn{1}{c}{}& \multicolumn{1}{c}{$\bullet$}& \multicolumn{1}{c}{$\bullet$} &  \multicolumn{1}{c}{AV-000}\\ 
    \multicolumn{1}{p{5.4cm}}{Build dependencies from sources} & \multicolumn{1}{c}{} & \multicolumn{1}{c}{\checkmark} & \multicolumn{1}{c}{} & \multicolumn{1}{c}{}& \multicolumn{1}{c}{}& \multicolumn{1}{c}{$\bullet$}& \multicolumn{1}{c}{$\bullet$} &  \multicolumn{1}{c}{AV-400, AV-500}\\ 
    \multicolumn{1}{p{5.4cm}}{Typo guard/Typo detection} & \multicolumn{1}{c}{} & \multicolumn{1}{c}{\checkmark} & \multicolumn{1}{c}{\checkmark} & \multicolumn{1}{c}{}& \multicolumn{1}{c}{}& \multicolumn{1}{c}{$\bullet$}& \multicolumn{1}{c}{$\bullet$} &  \multicolumn{1}{c}{AV-200}\\ 
    \multicolumn{1}{p{5.4cm}}{Establish vetting process for Open-Source components hosted in internal/public repositories} & \multicolumn{1}{c}{} & \multicolumn{1}{c}{\checkmark} & \multicolumn{1}{c}{} & \multicolumn{1}{c}{}& \multicolumn{1}{c}{}& \multicolumn{1}{c}{$\bullet$}& \multicolumn{1}{c}{$\bullet$} &  \multicolumn{1}{c}{AV-000}\\ 


    \multicolumn{1}{p{5.4cm}}{Runtime Application Self-Protection (RASP)} & \multicolumn{1}{c}{} & \multicolumn{1}{c}{} & \multicolumn{1}{c}{\checkmark} & \multicolumn{1}{c}{\checkmark}& \multicolumn{1}{c}{}& \multicolumn{1}{c}{}& \multicolumn{1}{c}{$\bullet$} &  \multicolumn{1}{c}{AV-000}\\ 
    \multicolumn{1}{p{5.4cm}}{Remove un-used dependencies} & \multicolumn{1}{c}{} & \multicolumn{1}{c}{\checkmark} & \multicolumn{1}{c}{} & \multicolumn{1}{c}{} & \multicolumn{1}{c}{}& \multicolumn{1}{c}{}& \multicolumn{1}{c}{$\bullet$}&  \multicolumn{1}{c}{AV-001}\\ 
    \multicolumn{1}{p{5.4cm}}{Prevent script execution} & \multicolumn{1}{c}{} & \multicolumn{1}{c}{\checkmark} & \multicolumn{1}{c}{} & \multicolumn{1}{c}{} & \multicolumn{1}{c}{}& \multicolumn{1}{c}{}& \multicolumn{1}{c}{$\bullet$}&  \multicolumn{1}{c}{AV-000}\\ 
    \multicolumn{1}{p{5.4cm}}{Code isolation and sandboxing} & \multicolumn{1}{c}{} & \multicolumn{1}{c}{} & \multicolumn{1}{c}{} & \multicolumn{1}{c}{\checkmark} & \multicolumn{1}{c}{}& \multicolumn{1}{c}{}& \multicolumn{1}{c}{$\bullet$}&  \multicolumn{1}{c}{AV-000}\\ 
    \multicolumn{1}{p{5.4cm}}{Version pinning} & \multicolumn{1}{c}{} & \multicolumn{1}{c}{\checkmark} & \multicolumn{1}{c}{} & \multicolumn{1}{c}{} & \multicolumn{1}{c}{}& \multicolumn{1}{c}{}& \multicolumn{1}{c}{$\bullet$}&  \multicolumn{1}{c}{AV-001}\\ 
    \multicolumn{1}{p{5.4cm}}{Dependency resolution rules} & \multicolumn{1}{c}{} & \multicolumn{1}{c}{\checkmark} & \multicolumn{1}{c}{} & \multicolumn{1}{c}{}& \multicolumn{1}{c}{}& \multicolumn{1}{c}{}& \multicolumn{1}{c}{$\bullet$} &  \multicolumn{1}{c}{AV-501, AV-508, AV-509}\\ 
    \multicolumn{1}{p{5.4cm}}{Establish internal repository mirrors and reference one private feed, not multiple} & \multicolumn{1}{c}{} & \multicolumn{1}{c}{\checkmark} & \multicolumn{1}{c}{} & \multicolumn{1}{c}{} & \multicolumn{1}{c}{}& \multicolumn{1}{c}{}& \multicolumn{1}{c}{$\bullet$}&  \multicolumn{1}{c}{AV-501,AV-502, AV-504, AV-505}\\ 
    \multicolumn{1}{p{5.4cm}}{Integrate Open-Source vulnerability scanner into CI/CD pipeline} & \multicolumn{1}{c}{} & \multicolumn{1}{c}{} & \multicolumn{1}{c}{\checkmark} & \multicolumn{1}{c}{} & \multicolumn{1}{c}{}& \multicolumn{1}{c}{}& \multicolumn{1}{c}{$\bullet$}& \multicolumn{1}{c}{AV-000} \\ 
    \multicolumn{1}{p{5.4cm}}{Integrity check of dependencies through cryptographic hashes} & \multicolumn{1}{c}{} & \multicolumn{1}{c}{} & \multicolumn{1}{c}{\checkmark} & \multicolumn{1}{c}{}& \multicolumn{1}{c}{}& \multicolumn{1}{c}{}& \multicolumn{1}{c}{$\bullet$} &  \multicolumn{1}{c}{AV-400, AV-500}\\ 
    
    \bottomrule
    \end{tabular}
    \caption{Safeguards against \ac{OSS} supply chain attacks, incl. control
    type, stakeholder(s) involved in their implementation, and a mapping to
    mitigated attack vectors (cf. Figure~\ref{fig:attacktreetaxonomy} to resolve their
    identifiers).}
    \label{tab:safeguards}
    \end{table*}

\section{Scientific and Grey Literature Resources}\label{appendix:C}

The four digital libraries queried during the \ac{SLR} are:
Google Scholar\footnote{\url{https://scholar.google.com/}}, arXiv\footnote{\url{https://arxiv.org/}},
IEEExplore\footnote{\url{https://ieeexplore.ieee.org/}} and ACM Digital Library\footnote{\url{https://dl.acm.org/}}.
The main sources used during the grey literature review are the following:
\begin{itemize}
    \item IQT Lab's Software Supply Chain Compromises dataset~\footnote{\url{https://github.com/IQTLabs/software-supply-chain-compromises}};
    \item Backstabber's Knife Dataset~\footnote{\url{https://dasfreak.github.io/Backstabbers-Knife-Collection/}}~\cite{ohm2020backstabbers};
    \item Whitepapers from Microsoft~\cite{microsoftwhitepaper} and Google~\cite{googlewhitepaper2};
    \item Whitepapers of projects for securing the software supply chain, like \ac{SLSA}~\cite{slsaframework}, sigstore~\footnote{\url{https://www.sigstore.dev/}}, TUF~\footnote{\url{https://theupdateframework.io}}, in-toto~\footnote{\url{https://in-toto.io}} and OSSF Scorecard~\footnote{\url{https://github.com/ossf/scorecard}};
    \item News aggregator (e.g., The Hacker News~\footnote{\url{https://thehackernews.com}}, Bleepingcomputer~\footnote{\url{https://www.bleepingcomputer.com}}, heise Security~\footnote{\url{hhttps://www.heise.de/security/}});
    \item Blogs of package repositories and security vendors (e.g., Snyk~\footnote{\url{https://snyk.io/blog/}}, Sonatype~\footnote{\url{https://blog.sonatype.com}}) and security researchers;
    \item Keynotes from cyber-security conferences (e.g., Blackhat~\cite{almubayed2019practical});
    \item MITRE's \ac{CAPEC}\footnote{\url{https://capec.mitre.org/}};
    \item MITRE's ATT\&CK\footnote{\url{https://attack.mitre.org}}.
\end{itemize}

\end{appendices}
    
\begin{acronym}[TDMA]
    \acro{CTI}{Cyber Threat Intelligence}
    \acro{TUF}{The Update Framework}
    \acro{PKI}{Public Key Infrastructure}
    \acro{CI}{Continuous Integration}
    \acro{CD}{Continuous Delivery}
    \acro{UI}{User Interface}
    \acro{VCS}{Versioning Control System}
    \acro{VMs}{Virtual Machines}
    \acro{VA}{Vulnerability Assessment}
    \acro{VCS}{Version Control System}
    \acro{SCM}{Source Control Management}
    \acro{IAM}{Identity Access Management}
    \acro{CDN}{Content Delivery Network}
    \acro{UX}{User eXperience}
    \acro{SLR}{Systematic Literature Review}
    \acro{SE}{Social Engineering}
    \acro{MITM}{Man-In-The-Middle}
    \acro{SBOM}{Software Bill of Materials}
    \acro{MFA}{Multi-Factor Authentication}
    \acro{AST}{Application Security Testing}
    \acro{RASP}{Runtime Application Self-Protection}
    \acro{OSS}{Open-Source Software}
    \acro{TARA}{Threat Assessment and Remediation Analysis}
    \acro{CAPEC}{Common Attack Pattern Enumeration and Classification}
    \acro{DoS}{Denial of Service}
    \acro{SCA}{Software Composition Analysis}
    \acro{SLSA}{Supply-chain Levels for Software Artifacts}
    \acro{SDLC}{Software Development Life-Cycle}
    \acro{ICT}{Information and Communication Technologies}
    \acro{C-SCRM}{Cyber Supply Chain Risk Management}
    \acro{DDC}{Diverse Double-Compiling}
    \acro{OSINT}{Open Source Intelligence}
    \acro{U/C}{Utility-to-Cost}
\end{acronym}

\end{document}